\documentclass[pra,twocolumn,superscriptaddress,footinbib,showpacs]{revtex4}

\usepackage{graphicx}
\usepackage{epsf}

\newcommand{\bq}{\begin{equation}}
\newcommand{\eequ}{\end{equation}}
\newcommand{\bqa}{\begin{eqnarray}}
\newcommand{\eqa}{\end{eqnarray}}

\newcommand{\ra}{\rangle}
\newcommand{\ket}[1]{\ensuremath{| {#1} \ra}}

\def\gt{\ensuremath{\tilde g}}
\def\kt{\ensuremath{\tilde k}}
\def\kapt{\ensuremath{\tilde \kappa}}
\def\gamt{\ensuremath{\tilde \gamma}}
\def\ut{\ensuremath{\tilde u}}
\def\Delt{\ensuremath{\tilde \Delta}}
\def\expct#1{\ensuremath{\langle #1\rangle}}

\def\Xe{\ensuremath{\expct{X}_\mathrm{e}}}
\def\Pe{\ensuremath{\expct{P}_\mathrm{e}}}
\def\dXe{\ensuremath{d\expct{X}_\mathrm{e}}}
\def\dPe{\ensuremath{d\expct{P}_\mathrm{e}}}
\def\Vxe{\ensuremath{{V_X^\mathrm{e}}}}
\def\Vpe{\ensuremath{{V_P^\mathrm{e}}}}
\def\dVxe{\ensuremath{dV_X^\mathrm{e}}}
\def\dVpe{\ensuremath{dV_P^\mathrm{e}}}
\def\Ce{\ensuremath{C_\mathrm{e}}}
\def\dCe{\ensuremath{dC_\mathrm{e}}}
\def\dWe{\ensuremath{dW_\mathrm{e}}}
\def\Ae{\ensuremath{A_\mathrm{e}}}

\def\Veff{\ensuremath{V_\mathrm{eff}}}
\def\Heff{\ensuremath{H_\mathrm{eff}}}
\def\capsubsz{\scriptscriptstyle}
\def\omegaL{\ensuremath{\omega_\mathrm{\capsubsz L}}}
\def\omegaA{\ensuremath{\omega_\mathrm{\capsubsz A}}}
\def\omegaHO{\ensuremath{\omega_\mathrm{\capsubsz HO}}}
\def\prty{\ensuremath{\mathcal{P}}}
\def\reduceop{\ensuremath{\mathcal{R}}}
\def\prtyr{\ensuremath{\mathcal{R^\dagger PR}}}

\def\psiket{\ensuremath{\ket{\psi}}}
\def\Vmax{\ensuremath{{V_\mathrm{max}}}}
\def\twoktDeltaXe{2\kt\Delta\!\Xe}
\def\DtG{\ensuremath{\Delta t_\mathrm{\capsubsz G}}}
\def\taud{\ensuremath{\tau_\mathrm{d}}}

\begin{document}
\title{Feedback cooling of atomic motion in cavity QED
                    \vbox to 0pt{\vss
                    \hbox to 0pt{\hskip-37pt\rm LA-UR-04-4687\hss}
                    \vskip 25pt}}

\author{Daniel A. Steck}
\affiliation{Theoretical Division (T-8), MS B285, Los Alamos National
  Laboratory, Los Alamos, NM 87545}
\affiliation{Oregon Center for Optics and Department of Physics, 1274 University 
of Oregon, Eugene, OR 97403-1274}
\author{Kurt Jacobs} 
\affiliation{Theoretical Division (T-8), MS B285, Los Alamos National
  Laboratory, Los Alamos, NM 87545}
\affiliation{Centre for Quantum Computer Technology, 
Centre for Quantum Dynamics, School of Science,
Griffith University, Nathan 4111, Australia}
\affiliation{Quantum Science and Technologies Group, Hearne Institute for 
Theoretical Physics, Department of Physics and Astronomy, Louisiana State 
University, 202 Nicholson Hall, Tower Dr., Baton Rouge, LA 70803}
\author{Hideo Mabuchi}
\affiliation{Norman Bridge Laboratory of Physics 12-33,
California Institute of Technology, Pasadena, CA 91125}
\author{Salman Habib} 
\affiliation{Theoretical Division (T-8), MS B285, Los Alamos National
  Laboratory, Los Alamos, NM 87545}
\author{Tanmoy Bhattacharya}
\affiliation{Theoretical Division (T-8), MS B285, Los Alamos National
  Laboratory, Los Alamos, NM 87545}

\begin{abstract}
We consider the problem of controlling the motion of an atom trapped 
in an optical cavity using continuous feedback. In order to realize 
such a scheme  experimentally, one must be able to perform state 
estimation of the atomic motion in real time. While in theory this 
estimate may be provided by a stochastic master equation describing 
the full dynamics of the observed system, integrating this equation 
in real time is impractical. Here we  derive an approximate 
estimation equation for this purpose, and use it 
as a drive in a feedback algorithm
designed to cool the motion of the atom. We examine the 
effectiveness of such a procedure using full simulations of the 
cavity QED system, including the quantized motion of the atom in one 
dimension.
\end{abstract}
\maketitle

\section{Introduction}
The theory of quantum measurement has been mired in controversy for 
the majority of its history. In a large part this is due to the 
philosophical difficulties, usually referred to as the ``quantum 
measurement problem,'' with which it is associated~\cite{Wheeler83}. However, 
in most areas of physics it has been generally possible to ignore 
quantum measurement theory, because it was largely irrelevant to real 
measurements that were being made by experimentalists~\cite{Milburn94}. In 
particular, if needed at all, an ensemble picture in which quantum 
measurement theory merely predicted average statistics provided a 
sufficient description. However, experimental technology has now 
advanced to the point where repeated measurements may be made on a 
\textit{single} quantum system as it evolves~\cite{Mabuchi99, 
Muenstermann99, Hood00, Pinkse00}. As a 
result, not only does quantum measurement theory become directly 
relevant to experimental physics, but we can consider testing the 
predictions of this theory regarding the effects of measurement on 
the single system itself.

While the theory of quantum measurement is now widely accepted, some 
still remain unconvinced that the conditioned state 
describes what is really happening to an \textit{individual} system when 
it is measured. How might we verify these predictions, and bring at 
least this part of the controversy to an end? One way to do this is 
to attempt to perform feedback control on an individual quantum 
system \cite{Belavkin83,Belavkin87}.
Such an experiment involves taking into account the changes in the 
system due to the measurement results, and using this knowledge to 
alter inputs to the system in order to control its dynamics. 
This would therefore provide a direct verification of quantum measurement 
theory: if 
the feedback algorithm works, then the theory is correctly predicting 
the behavior of the system; conversely, if the theory is not giving 
accurate predictions, the feedback algorithm will be ineffective.

Cavity quantum electrodynamics (CQED) is one area in 
which the motion of a single quantum system, an 
individual atom in an optical cavity, can be monitored experimentally in real 
time~\cite{Mabuchi99, Muenstermann99, Hood00, Pinkse00, Sauer04}. 
Not only does the light emitted from the 
optical cavity provide a means to track the atomic motion, but the 
laser driving the cavity provides a means to apply forces to the 
atom. This system is therefore an excellent candidate for implementing 
real-time quantum feedback control. It is our purpose here to examine the 
implementation of such a process in this system, and to show that by 
using approximate estimation techniques, which will be essential in 
such experiments for the foreseeable future,
a measurable degree of control over the atomic 
motion can be realized. In particular, we will simulate a control
algorithm designed to cool the atomic motion. This might be referred 
to as ``active'' cooling (control)
\cite{Belavkin83, Raizen98, Belavkin99, Doherty99, Doherty00, Armen02, Fischer02, Vitali03, Hopkins03, Geremia03, Geremia04, vanHandel04, Geremia04b, Geremia05, Smith04, Smith02, James04, Reiner04, Hanssen02, Morrow02, Vuletic04, Steck04,  Rabl05, Steixner05, Wiseman05}, 
as opposed to passive cooling schemes which 
have also been the focus of recent theoretical 
\cite{Horak97, Alge97, Gangl00, Gangl00b, Vuletic00, Domokos01, Gangl01, Horak01, vanEnk01}
and experimental \cite{Maunz04} work.

In general, feedback control of the atomic motional
state requires the real-time estimation of the quantum state from the 
continuous measurement, thus providing the information needed to decide 
how the system should be perturbed to bring it to the target state.
The theoretical tools required for obtaining this continuous estimate are 
those of continuous quantum measurement~\cite{Belavkin87, Belavkin88, 
Belavkin89, Chruseinski92, Srinivas81, Barchielli83, Barchielli85, Diosi86, Diosi88, Gisin84, Caves87, Barchielli93, Carmichael93, Wiseman93}. 
An estimate of the state of the quantum system (in our case the motional 
state of the atom), conditioned continuously upon the results of 
measurement, is given by a stochastic master equation (SME). 
The concept of an SME was first developed by Belavkin in~\cite{Belavkin87} 
(as the quantum equivalent of the classical Kushner-Stratonovich equation), 
and a derivation of an SME in the language of quantum optics may 
be found in~\cite{Wiseman93}.
In principle the observer may calculate this state estimate by 
integrating the relevant SME using the measurement results, and from 
this determine the values of the inputs to the system to effect 
control. 
However, in practice the complexity of the SME will prevent 
the observer from calculating the  state estimate in real time 
for systems where the dynamics are sufficiently fast, such as the 
atomic dynamics we consider here. As a result, a simplified 
state-estimation procedure is essential.

The theory of feedback control for classically observed linear 
systems is well developed~\cite{Jacobs93, Maybeck82, Whittle96, Zhou95}. 
This theory considers the optimality and robustness of feedback 
algorithms for given resources, and in some special cases may be 
applied directly to the control of quantum systems~\cite{Belavkin87, 
Doherty99}. 
Discussions of the application of classical feedback control 
techniques to quantum systems may be found in Refs.~\cite{Belavkin99, Doherty99, Doherty00}. 
However, for nonlinear control problems \cite{Doherty00, Wiseman02, 
Jacobs03},
such as that of the atomic 
motion we consider, no general results regarding the optimality of control 
algorithms exist as of yet. 
Here we will be concerned 
with presenting simple control algorithms for cooling the atoms,
and demonstrating 
that along with an approximate estimation algorithm, a substantial 
intervention into the dynamics of the atom using real-time feedback 
control should be achievable experimentally in the near future. 
In doing so, we build upon previous results
\cite{Steck04} by providing a detailed analysis of the
cooling performance in this system, including its robustness
to parameter variations and real-world limitations on computational
speed and detection efficiency.  Furthermore, we are able to
anticipate several difficulties in an experimental implementation
and suggest solutions to overcome them.

Owing to the computationally expensive nature of simulating an atom 
in an optical cavity, including the atomic motion in one dimension, 
supercomputers are invaluable for the task of evaluating
the performance of the control algorithm over many realizations. This 
is the first time that full quantum simulations of such a CQED
system, including the quantized, one-dimensional motion of the atom, 
have been performed. We are thus able to confirm the results 
of various approximations which have been used previously to provide 
a simplified analysis of this system~\cite{Holland91, Storey92, Quadt95, 
Doherty98, Doherty99}.

\section{Description of the system}\label{thesys}
The system we consider here is that of a two-level atom in a driven, 
single-mode optical cavity, where the output from the cavity is 
monitored using homodyne detection~\cite{Carmichael93, Wiseman93}. Due to the 
continuous observation, the evolution of the system is described by a 
stochastic master equation (SME) for the density operator, which 
can be written in It\^o form as~\cite{noteSME}
\begin{equation}
  \setlength{\arraycolsep}{0ex}
  \renewcommand{\arraystretch}{1.8}
  \begin{array}{rcl}
    d\rho &{}={}& \displaystyle -i[(H/\hbar),\rho]dt 
              + \gamma dt \int_{-\hbar k}^{\hbar k} N(u) 
                   \mathcal{D}[\sigma e^{-iux/\hbar}]\rho \, du \\
        && \displaystyle + \kappa \mathcal{D}[a]\rho\, dt 
           + \sqrt{\eta\kappa} \mathcal{H}[a]\rho\, dW ,
  \end{array}
  \label{SME}
\end{equation}
where 
the Hamiltonian is~\cite{Doherty98}
\begin{equation}
  \frac{H}{\hbar} = \frac{p^2}{2m\hbar} + g\cos(kx) (\sigma^\dagger 
  a e^{i\Delta t} + \mathrm{H.c.})  - E (a+a^\dagger) ,
\end{equation}
and we have defined the superoperators $\mathcal{D}[c]$ and $\mathcal{H}[c]$
by
\begin{equation}
  \setlength{\arraycolsep}{0ex}
  \renewcommand{\arraystretch}{1.8}
  \begin{array}{rcl}
    \mathcal{D}[c]\rho &{}:={}& c\rho c^\dagger - \frac{1}{2}(c^\dagger c \rho
      + \rho c^\dagger c)\\
    \mathcal{H}[c]\rho &{}:={}& c\rho +\rho c^\dagger - \expct{c+c^\dagger}\rho 
  \end{array}
\end{equation}
for an arbitrary operator $c$. 
In addition, the measured homodyne signal, given by 
\begin{equation}
   dr(t) = \eta\kappa \langle a + a^\dagger \rangle dt 
               + \sqrt{\eta{\kappa}} \, dW, 
   \label{mr}
\end{equation}
continuously provides information about the optical phase shift due to the
atom-cavity system and hence, as we will see below, information about 
the atomic position.
In the above equations, $x$ is an 
operator representing the atomic position along the length of the 
cavity, $p$ is the conjugate momentum operator, $a$ is the 
annihilation operator for the cavity mode, and $\sigma$ is the 
lowering operator for the atomic internal states. The energy loss rate of 
the cavity (or equivalently, the full-width-at-half-maximum cavity transmission)
is $\kappa$, and $E$ gives the strength of the driving 
laser, and is related to the laser power $P$ by $E = \sqrt{\kappa 
P/(\hbar\omegaL)}$, where $\omegaL$ is the angular frequency of the 
laser light. We take the laser frequency to be resonant with the 
cavity mode. The laser angular wave number is given by 
$k=\omegaL/c$, the decay rate of the atomic excited state is denoted 
by $\gamma$, $g$ is the CQED coupling constant between the atomic 
internal states and the cavity mode, and $\Delta$ is the detuning 
between the atom and the cavity mode, given by 
$\Delta=-(\omegaL-\omegaA)$, where $\omegaA$ is the atomic transition frequency. 
We will assume a positive (red) detuning, corresponding to a 
trapping optical potential.
Also, $N(u)$ is the probability that the atomic momentum recoil along the
$x$-direction is $u$ due to a spontaneous emission event, and
$dW$ is the differential of a stochastic Wiener process.
Note that we 
have written the stochastic master equation in the interaction 
picture where the free oscillation of the light and 
the dipole rotation of the atom are implicit. 

The density operator $\rho(t)$ obtained as a solution to the SME 
condenses our knowledge of the initial condition and the 
measurement record, allowing a prediction of the statistics of any 
future measurement on the system.  As we show numerically in 
Section~\ref{sec:estimation-cooling-dynamics}, the solution 
to the equation is, generically, insensitive 
to the initial conditions except at very early times.  
We therefore refer to the solution of the SME as the 
observer's \textit{best estimate} or \textit{state of knowledge} of the quantum system.

In order to perform numerical calculations, it is sensible to choose 
units for position, momentum and time that are relevant for the corresponding
scales of the system. When the detuning of the atom from the cavity 
is large compared to the cavity decay rate and the CQED coupling 
constant, both the internal atomic states and the cavity mode may be 
adiabatically eliminated, and in this case the atom sees an effective 
potential, given by
\begin{equation}
  \Veff(x) = -\hbar \frac{g^2\alpha^2}{\Delta}\cos^2(kx), 
\end{equation}
where $\alpha := 2E/\kappa$. Since this regime is the one in which we will 
be primarily interested, convenient units for $x$ and $p$ are 
given respectively by the width in position and momentum of the 
ground-state wave function in the harmonic approximation to one of the 
cosine wells. A convenient unit for time is set by the frequency of 
oscillation in the same harmonic potential. Thus, given the three 
scales of the problem,
\begin{equation}
  \begin{array}{rcl}
     \omegaHO &{}:={}& \alpha g k \sqrt{2\hbar/|m\Delta|}  \\
     w_x &{}:={}& \sqrt{\hbar/(m\omegaHO)} \\
     w_p &{}:={}& \sqrt{\hbar m\omegaHO} ,
  \end{array}
\end{equation}
we will use the 
scaled variables $X := x/w_x$ and $P := p/w_p$, and we will measure time
in units of $2\pi/\omegaHO$. 
The SME for the density matrix may now be written 
as
\begin{equation}
  \setlength{\arraycolsep}{0ex}
  \renewcommand{\arraystretch}{1.8}
  \begin{array}{rcl}
    d\rho &{}={}& -i[H,\rho]dt + \tilde{\gamma} \,dt
                  \int_{-\tilde{k}}^{\tilde{k}} N(\tilde{u}) 
	   	\mathcal{D}(\sigma e^{-i\tilde{u}X})\rho \, d\tilde{u} \\
        &&+ \tilde{\kappa} \mathcal{D}[a]\rho\, dt + 
            \sqrt{\eta\tilde{\kappa}} \mathcal{H}[a]\, dW ,
  \label{scSME}
  \end{array}
\end{equation}
where
\begin{equation}
  H = \pi P^2  + \tilde{g} \cos(\tilde{k}X) 
(\sigma^\dagger a e^{i\tilde{\Delta} t} + \mbox{H.c}) - \tilde{E} 
(a+a^\dagger) 
\end{equation}
and
\begin{equation}
   d\tilde{r}(t) = \eta \langle a + a^\dagger \rangle dt 
               + \sqrt{\eta} \, dW, 
   \label{mrscaled}
\end{equation}
with $d\tilde{r}(t) = dr(t)/\sqrt{\kapt}$.
The scaled rate constants $\tilde{\kappa}, \tilde{E}, \tilde{\Delta}, 
\tilde{\gamma}$, and $\tilde g = \sqrt{\pi\Delt}/(\alpha\kt)$ 
are obtained from 
their absolute counterparts by dividing by $\omegaHO/(2\pi)$. The scaled 
wave number is naturally $\tilde{k} = k w_x$. Note that in these 
scaled units we have scaled $\hbar$ out of the problem, so that $[X,P]=i$.

While we will solve this equation numerically, it is nevertheless 
useful to examine the solutions obtained by adiabatically eliminating 
both the atomic internal states and the cavity mode, 
which is possible when the detuning  
$\tilde{\Delta}$ is much larger than the coupling constant 
$\tilde{g}$ so that the excited atomic state is nearly unpopulated during 
the evolution \cite{Doherty98}. In 
this case the Hamiltonian part of the evolution is governed by the 
effective Hamiltonian 
\begin{equation}
   \Heff := \pi P^2 - \frac{\tilde{g}^2}{\tilde{\Delta}}                                            
   \cos^2(\tilde{k}X) a^\dagger a .
\end{equation} 
Thus, the atom moves in a sinusoidal potential, with period 
$\pi/\tilde{k}$. The average height of the potential wells 
is essentially proportional
to $\langle a^\dagger a \rangle$. It is 
therefore instructive to consider the cavity-mode dynamics, 
which determine the magnitude of the effective potential. Under the 
effective Hamiltonian, the Heisenberg equations of motion for the 
relevant cavity mode operators are
\begin{equation}
  \setlength{\arraycolsep}{0ex}
  \renewcommand{\arraystretch}{1.8}
  \begin{array}{rcl}
    \partial_t{a} &{}={}& \displaystyle i\tilde{E} - \frac{\tilde{\kappa}}{2} a + 
    i\frac{\tilde{g}^2}{\tilde{\Delta}}\cos^2(\kt X) a + \sqrt{\tilde{\kappa}} 
    a_\mathrm{in}(t)  \\
    \partial_t{(a^\dagger a)} & = & \displaystyle 
     -\tilde{\kappa}a^\dagger a - i \tilde{E}(a 
    - a^\dagger) \\ &&\displaystyle 
    +\sqrt{\tilde{\kappa}}(a^\dagger a_\mathrm{in}(t) + 
    a_\mathrm{in}^\dagger(t) a) ,
  \end{array}
\end{equation}
where $a_{\mathrm{in}}(t)$ is the input quantum noise operator coming 
from the electromagnetic field outside the cavity, and satisfies 
$[a_{\mathrm{in}}(t),a_{\mathrm{in}}^\dagger(t+\tau)]=\delta(\tau)$. We can 
now identify a regime in which $\tilde{\kappa}$ is large compared to 
$\tilde{g}^2/\tilde{\Delta}$. In this case the cavity mode damps 
quickly to a steady state, and we can approximate the solution by
\begin{equation}
  \setlength{\arraycolsep}{0ex}
  \renewcommand{\arraystretch}{2.1}
  \begin{array}{rcl}
     a &{}\approx{}& \displaystyle
              i\alpha \left[ 1 -  i2\left(\frac{\tilde{g}^2}{\tilde{\kappa}\tilde{\Delta}}\right)
		   \cos^2(\tilde{k}X) \right]^{-1} ,  \\
       &{}\approx{}& \displaystyle i\alpha  - \alpha \left( 
              \frac{2\tilde{g}^2}{\tilde{\Delta}\tilde{\kappa}} 
              \right) \cos^2(\tilde{k}X) , \\
  a^\dagger a &{}\approx{}& \displaystyle\alpha \frac{(a - a^\dagger)}{2i} .
  \end{array}
\end{equation}
There will be fluctuations about these values due to the noise 
with a spectrum with a width of the order of $\tilde{\kappa}$. When 
$\kappa$ (or $\tilde{\kappa}$) is sufficiently large, 
then $\langle a^\dagger a \rangle \approx \alpha^2$, and the height 
of the potential will therefore be determined by the strength of the 
driving. As a consequence, one can manipulate the potential that the 
atom sees by changing the strength  $\tilde{E}$ of the driving. 
The measured homodyne signal is given in the adiabatic approximation by 
\begin{equation}
   d\tilde{r}(t) =  -
      \sqrt{8\eta^2\Gamma} \langle 
      \cos^2(\kt X)\rangle dt + \sqrt{\eta} \, dW ,
   \label{mradiab}
\end{equation}
where $d\tilde{r}(t):= dr(t)/\sqrt{\kapt}$; 
the position-information content of 
this signal is more clear in this form.

\section{Cooling the atomic motion using feedback}\label{alg}
As mentioned in the introduction, in implementing feedback control, 
two considerations are paramount. The first is the ability to effectively 
track the evolution of the system in real time (state estimation).
The second is to provide an algorithm which determines how this 
information about the system should be used to alter the inputs to 
the system to effect control. We now consider these two questions in 
turn.

\subsection{State Estimation in Real Time}\label{section:gest}
The SME given in Eq.~(\ref{SME})  continuously provides the observer's 
best estimate of the state of the system from the information 
provided by the measurement record $r(t)$ [Eq.~(\ref{mr})]. Therefore, 
it is possible, in principle, to use the measurement record to integrate the 
full SME to obtain a continuous state estimate. However, the 
resources required to do this in real time are prohibitive in practice. This is 
because the SME, being a partial stochastic differential equation, 
contains a large effective number of variables, and is almost
impossible to integrate in real time, given that the time scale of 
the atomic motion is typically a few $\mu$s. As a result, 
we must obtain an estimation equation which contains only a small 
number of variables, but is nevertheless a good approximation to the 
full SME.

To obtain a compact system of estimation equations, we first note that initially
Gaussian states remain close to Gaussian for a fairly large number of 
oscillations in a sufficiently strong sinusoidal potential. 
We thus make a 
Gaussian approximation for the density matrix, which amounts to ignoring 
all cumulants higher than second order. 
In the present system, which effectively measures
$\expct{\cos^2(\kt X)}$, we expect this approximation to remain valid as long
as the wave packet is much narrower than a single well ($\kt \Delta X\ll 1$), 
and the wave-packet momentum width
is much larger than two photon recoils ($\Delta P \gg 2\kt$). 
To further simplify the estimation equations, we consider 
the SME after adiabatic elimination of the cavity
and internal atomic degrees of freedom, given by
\begin{equation}
  \setlength{\arraycolsep}{0ex}
  \renewcommand{\arraystretch}{1.5}
  \begin{array}{rcl}
    d\rho &{}={}& \displaystyle -i[\Heff,\rho]\, dt
      + 2 \Gamma\mathcal{D}[\cos^2(\kt X)]\rho\, dt \\
      &&\displaystyle -\sqrt{2\eta\Gamma}\mathcal{H}[\cos^2(\kt X)]\rho\, dW,
  \end{array}
  \label{adiabaticSME}
\end{equation}
which gives a three-parameter model ($\kt$, $\Gamma$, and 
$\eta$).
Here, we have defined the effective Hamiltonian
\begin{equation}
  \Heff = \pi P^2 - \Vmax\cos^2(\kt X),
\end{equation}
where the maximum light shift 
\begin{equation}
   \Vmax := \frac{\alpha^2\gt^2}{\Delt} = \frac{\pi}{\kt^2}
\end{equation}
is not an independent parameter in this static problem, but we will
treat it as such when considering feedback via a time-dependent potential. 
We have also defined the effective measurement strength
\begin{equation}
  \Gamma := \frac{2\alpha^2 \tilde{g}^4}{\tilde{\Delta}^2\tilde{\kappa}}.
  \label{Gammadef}
\end{equation}
We further simplify the estimation model by making a Gaussian 
ansatz for the density operator. 
The resulting estimation equations contain merely 5 
variables: the estimated mean position and momentum, \Xe\ 
and \Pe, the estimated 
variances, \Vxe\ and \Vpe, and the estimated 
symmetrized covariance $\Ce := \langle XP + 
PX\rangle_{\mathrm{e}}/2 - \Xe\Pe$. 
(Due to the information content of the measurement process, the two 
variances and the covariance are algebraically independent, in
contrast to a closed system.)
Using the equation of motion for the expectation value of
an arbitrary operator $A$,
\begin{equation}
  \setlength{\arraycolsep}{0ex}
  \renewcommand{\arraystretch}{1.4}
  \begin{array}{rcl}
    d\expct{A}&{}={}&\mathrm{Tr}[A\,d\rho]\\ &{}={}&\displaystyle
      -i\expct{[A,\Heff]}dt
      + 2 \Gamma \expct{D[\cos^2(\kt X)]A}dt\\ && \displaystyle
      {}- \sqrt{2\eta\Gamma}\expct{\mathcal{H}[\cos^2(\kt X)]A}dW,
  \end{array}
\end{equation}
we can calculate the system of five Gaussian estimation equations,
which are
\begin{equation}
  \setlength{\arraycolsep}{0ex}
  \renewcommand{\arraystretch}{1.4}
  \begin{array}{rcl}
    \dXe &{}={}& 2\pi \Pe dt \\
       && + \sqrt{8\eta\Gamma} \kt \Vxe 
       \exp(-2\kt^2\Vxe) \sin(2\kt\Xe) \dWe 
    \\ 
    \dPe &{}={}& \displaystyle -
      \Vmax 
      \kt \exp(-2\kt^2\Vxe) \sin(2\kt\Xe) dt \\
      & & + \sqrt{8\eta\Gamma} \kt \Ce \exp(-2\kt^2\Vxe)\sin(2\kt\Xe) \dWe 
    \\ 
    \dVxe &{}={}& 4\pi\Ce dt \\
    & & -8\eta\Gamma \kt^2 \Vxe^2 \exp(-4\kt^2\Vxe) \sin^2(2\kt\Xe) dt\\ 
    & & +2 \sqrt{8\eta\Gamma} \kt^2 \Vxe^2\exp(-2\kt^2\Vxe) \cos(2\kt\Xe) \dWe 
    \\ 
    \dVpe &{}={}&\displaystyle 
       -4 \Vmax 
       \kt^2\Ce\exp(-2\kt^2\Vxe)\cos(2\kt\Xe) dt\\
    && + \Gamma \kt^2[1-\exp(-8\kt^2\Vxe)\cos(4\kt\Xe)] dt\\
    && - 8\eta\Gamma \kt^2 \Ce^2 \exp(-4\kt^2\Vxe)\sin^2(2\kt\Xe) dt\\
    && - \sqrt{2\eta\Gamma}\kt^2[1-4
           (\Ce^2+\kt^2\Vxe)\exp(-2\kt^2\Vxe)]\\
       &&\hspace{6mm}\times\cos(2\kt\Xe)\dWe
    \\ 
    \dCe &{}={}& \displaystyle
      2\pi\Vpe dt\\
    &&\displaystyle -2 \Vmax 
            \kt^2\Vxe\exp(-2\kt^2\Vxe)\cos(2\kt\Xe) dt\\
    && - 8\eta\Gamma \kt^2 \Vxe \Ce \exp(-4\kt^2\Vxe)\sin^2(2\kt\Xe) dt\\
    && +2\sqrt{8\eta\Gamma}\kt^2\Vxe\Ce\exp(-2\kt^2\Vxe)\cos(2\kt\Xe)\dWe .
  \end{array}
  \label{gest}
\end{equation}
Although we can eliminate \Vmax\ from these estimator equations
(since $\Vmax\kt^2=\pi$), we keep the optical potential explicit
here because both $\Vmax$ and $\Gamma$ will effectively
become time-dependent
quantities when we consider feedback control by modulating the
amplitude of the optical potential.
The reconstructed ``Wiener increment'' \dWe\ arises by making the Gaussian 
approximation to the photocurrent equation (\ref{mradiab}):
\begin{equation}
  \dWe = \frac{d\tilde{r}(t)}{\sqrt{\eta}} 
   + \sqrt{2\eta\Gamma}[1+\exp(-2\kt^2\Vxe)\cos(2\kt\Xe)]dt .
\end{equation}
These estimation equations resemble stochastic differential equations,
but it is important to recognize that \dWe\ does not have the same statistics
as the usual Wiener increment $dW$.  
In particular, the reconstructed increment is the usual Wiener 
increment with an extra deterministic component,
\begin{equation}
  \setlength{\arraycolsep}{0ex}
  \begin{array}{rcl}
    \dWe &{}={}& dW + \sqrt{2\eta\Gamma}\left[
           \exp(-2\kt^2\Vxe)\cos(2\kt\Xe)\right. \\&& 
            \hspace{40mm}{}- \displaystyle\left.
           \langle\cos(2\kt X)\rangle\right] dt,
  \end{array}
  \label{dWefromdW}
\end{equation}
which reflects the difference between the actual and estimated
quantum states.  This is the route by which the measurement
information is incorporated into the estimator evolution.
Thus, these equations cannot be treated by the usual high-order
numerical methods for stochastic differential equations, which assume 
that the deterministic and stochastic parts of the differential 
equation can be explicitly separated.

Another important issue to note is that because the estimator only has 
information about $\langle \cos^2(\tilde{k}X) \rangle$, it cannot 
tell us in which well the atom is located, 
and therefore, while feedback will allow us to cool an atom within a 
well, we cannot know within \textit{which} well the atom is confined. 
Moreover, the estimator cannot tell us on what side of a well the 
atom is located. 
Note that this degeneracy is not a consequence of the Gaussian
approximation, but also applies to an observer using the 
full SME for estimation.
As a consequence, the estimator will show us the 
motion of the atom up to a phase factor of $\pi$: the estimated state 
will either be in phase with the true motion, or completely out of phase,
since these two motions are mirror images of one another. Thus, for 
the feedback algorithm to be effective for every run, it is important 
that it works for both cases without change. The algorithms that we 
describe in the next section have this property.

It is worth noting that the insensitivity of the feedback algorithm 
to the side of the well the atom is on has a potential advantage. In 
simulating the dynamics of the atom, we find that if the initial 
energy of the atom is large enough for it to move from one well to 
another, then as it does so the wave function tends to split across 
the interwell barrier, with part going back down into the original well, 
and part going down into an adjacent well. In this case the wave function 
is no longer Gaussian, but consists of two or more ``humps."
However, even though the two humps are on opposite sides 
of their respective wells, they have approximately the same downward 
motion. Thus, even though the estimator is Gaussian, 
the fact that the humps are in different wells is
immaterial to the measurement of $\langle\cos^2(\kt X)\rangle$,
and the estimator effectively tracks the evolution of 
an ``equivalent'' quantum state localized within a single well in the
sense of generating the same measurement record as the fragmented state.
As a result, we can 
expect the Gaussian estimator to continue to work even though the 
atom is distributed across more than one well. 

In order to perform the continuous estimation (whether using 
the full SME or the above Gaussian approximate estimator) one chooses 
an initial state for the estimator that reflects the initial ignorance
of the true quantum state and integrates 
the estimator equations using the measurement record as this record 
is obtained. In our case, when the atom is dropped into the cavity we 
have very little idea of where it is. It is therefore sensible to 
choose as the initial state for the estimator a Gaussian broad enough 
to cover one side of a well, and centered somewhere on that side (it 
is not important where precisely). We only need to include one side of 
the well in our initial uncertainty, since as mentioned above, both 
sides are equivalent from the point of view of the estimator.
In practice, though, it is sufficient to pick a state with a much
smaller uncertainty product,
centered on one side of a potential well.  Even if this
estimator state
does not substantially overlap the actual quantum state, as we will 
see below, it will
converge to the ``true'' state in a time of order $1/\Gamma$ as the
measurement information is incorporated.

\subsection{An Effective Cooling Algorithm}\label{section:alg}

As discussed in Section~\ref{thesys}, it is possible to raise and 
lower the height of the potential by changing the input laser power. 
Thinking of the atom as a classical particle, one would expect to be 
able to cool the atom using the following simple algorithm: When the 
atom is moving toward the center of the well, 
it is being accelerated by the potential, and so we can lower the 
laser power to reduce this acceleration. On the other hand, when the 
atom is moving away from the center of the well, then the potential 
decelerates the atom, and in this case we can increase the laser 
power so as to increase the deceleration. 
Writing the optical field amplitude in the absence of 
feedback as $E$, we will denote the value of the field during 
times when it is ``switched high'' as $(1 + \varepsilon_1) E$ and the 
value when it is ``switched low'' as $(1 - \varepsilon_2) E$.
Thus the potential height will have the respective values
$(1 + \varepsilon_1)^2 V$ and $(1 - \varepsilon_2)^2 V$, where $V$ is the
unmodulated potential depth.
In this way, we repeatedly 
switch the potential between a high value and a low value, which 
both slows the atom and moves it in towards center of the well, 
reducing the total energy and hence cooling the atom. This algorithm 
also has the nice property that all the information it requires is 
whether the atom is climbing or descending the well. Therefore,
the question of whether the estimated state is in phase or completely 
out of phase with the true motion is immaterial, since in both cases the relevant 
information is known. However, it turns out that this algorithm is not 
very effective at cooling the atomic motion, because while it does 
cool the centroid, it simultaneously feeds energy into 
the motion by squeezing the atomic wave function. A detailed discussion 
of this effect is given in Appendix A. The result is that this simple 
centroid-only cooling algorithm is not adequate to cool the atom to the 
ground state. To do so we must develop a more sophisticated algorithm 
which can reduce the motional energy associated with the variances of 
the wave function. 

To proceed, we consider the change in the motional energy due to time 
dependence of the optical-potential amplitude:
\begin{equation}
  \partial_t\langle {E}_\mathrm{eff}\rangle_\mathrm{fb}
    = -(\partial_t \Vmax)\langle\cos^2(\kt X)\rangle.
  \label{coolingenergy}
\end{equation}
It is clear from this expression that the ``bang-bang''
feedback strategy from the simple cooling algorithm, where the
potential amplitude is switched cyclically and 
suddenly between two values, 
maximizes
the energy extraction rate if we consider only algorithms
with a cyclic modulation of the potential within that range.
In fact, we should switch the potential low when the 
value of $-\langle\cos^2(\kt X)\rangle$ is maximized (i.e., the
modulus is minimized), thus
maximizing the energy extracted from the atom. Then we should switch
the potential high when $-\langle\cos^2(\kt X)\rangle$
is minimized, thus minimizing the energy transferred to the
atom when completing the modulation cycle.

When we must rely on the Gaussian estimator
to determine when to switch the potential, we switch the potential
low or high when the estimated quantity
\begin{equation}
  y_\mathrm{est} := -\exp(-2\kt^2 \Vxe)\cos(2\kt \Xe)
  \label{gestswitch}
\end{equation}
is maximized or minimized, respectively.  Doing so involves an
additional technical complication, however, since at any given
instant, we must predict whether or not $y_\mathrm{est}$ is
an extremal value.  To do this we dynamically fit a quadratic
curve to a history of these values (typically the last several
hundred), and then trigger the feedback transitions 
on the slope of the fitted curve.  This procedure is a convenient 
method for predicting the times when $y_\mathrm{est}$ is 
maximized or minimized, and helps to reject the residual noise in
the time evolution of this quantity.
More formally, given a 
set of estimates $y_{\mathrm{est}, n}$ corresponding to 
times $t_n$, we implement the curve fit of the
function $a_0+a_1 x + a_2 x^2$ to the last $q$ values of
$y_{\mathrm{est}, i}$.  We will defer the discussion of the
computational efficiency of this algorithm until Section 
\ref{section:speed}.

It may seem strange that this algorithm makes use of past state 
information, when all the necessary information should in principle
be contained in the current estimate of the state.  Indeed, it is 
clear from the
quantum Bellman equation \cite{Belavkin99, Belavkin87, Doherty00} that 
the optimal control algorithm is given by only looking forward
in time given the current state.  
In practice, such an algorithm is difficult to realize,
and the present algorithm is likely suboptimal but robust and
effective.

We also note that since we are triggering in this algorithm
on the estimated value of $-\langle\cos^2(\kt X)\rangle$, it
seems that it could be much simpler to simply trigger on the
detector photocurrent $d\tilde{r}(t)$, 
which from Eq.~(\ref{mradiab}) we know is
a direct measure of this quantity plus noise.  In fact, 
we can use the same curve-fitting procedure on $d\tilde{r}(t)$ to reject
the noise on the signal.  However, as we will see, this procedure
does not work nearly as well as with the Gaussian estimator, because
the signal $d\tilde{r}(t)$ is noise-dominated, and the Gaussian estimator
acts as a nearly optimal filter for the useful information from
this signal, in the same sense as a Kalman filter 
\cite{Maybeck82, Belavkin79, Doherty99}.

\subsection{Cooling Limits}\label{section:limits}

At this point, we can work out a simple theory of cooling for the
improved cooling algorithm developed 
in the previous section.  When we switch
the optical potential low, the
change in energy from Eq.~(\ref{coolingenergy}) is given by
\begin{equation}
  \Delta\langle {E}_\mathrm{eff}\rangle_\mathrm{fb}
    = -4\varepsilon \Vmax\langle\cos^2(\kt X)\rangle,
  \label{downtrans}
\end{equation}
where we have taken $\varepsilon_1=\varepsilon_2=\varepsilon$, so that the
potential amplitude is switched from $(1+\varepsilon)^2\Vmax$ to
$(1-\varepsilon)^2\Vmax$.  When we switch high, the expression
for the energy change is
the same except for an overall minus sign.

Now we consider these expectation values for a particle very close to the
ground state, and we will first consider the case where
all the excess motional energy is associated with the motion of 
the wave-packet centroid.
For the ground state itself, an equal part $E_0/2$ 
of the ground-state energy $E_0$ is associated with each of
the kinetic and potential parts of the effective Hamiltonian.
If we define $\Delta E := \langle {E}_\mathrm{eff}\rangle - E_0$
as the small, additional motional energy above the ground state,
then as the displaced wave packet (coherent state) 
evolves in the nearly harmonic
potential, 
the value $\Vmax\langle[1-\cos^2(\kt X)]\rangle$ of the 
potential part of the atomic energy oscillates between
$E_0/2$ and $E_0/2 + \Delta E$.
Then the best cooling that can be achieved for this state corresponds
to two up and two down transitions per oscillation period (which is
unity in our scaled units).  Inserting these extremal values of
$-\Vmax\langle\cos^2(\kt X)\rangle$ into Eq.~(\ref{downtrans})
and the corresponding expression for upward transitions 
gives the cooling rate of
\begin{equation}
  \Delta\langle {E}_\mathrm{eff}\rangle_\mathrm{fb}
    = -8\varepsilon(\langle {E}_\mathrm{eff}\rangle - E_0)
   \label{coolingrate}
\end{equation}
per unit time.  

In practice, the cooling algorithm switches more often than in this
estimate due to noise on the signal.
However, in such cases we might still expect that Eq.~(\ref{coolingrate}) is 
a reasonable estimate for the cooling rate. 
Even though the potential is switched more often, the switches are 
caused by noise and thus do not occur at optimal moments in time.
Thus the amount of cooling per switching cycle is reduced, and we
observe that the net cooling effect per unit time is near the optimal
rate of Eq.~(\ref{coolingrate}), presumably 
because the cooling algorithm is not too sensitive 
to the noise.
If we then assume that heating due to spontaneous emission is 
negligible compared to measurement heating, then the steady-state
motional energy is obtained when the cooling rate
in Eq.~(\ref{coolingrate}) balances the heating rate which we derive in 
Appendix~\ref{heating} (Eq.~(\ref{measurementheating})).
This gives the steady-state energy
\begin{equation}
  \langle {E}_\mathrm{eff}\rangle_\mathrm{ss}
    = \frac{E_0}{1-\beta},
   \label{E0ss}
\end{equation}
where $\beta := \Gamma\kt^4/2\varepsilon$.

The other extreme case we will consider is where the motional energy 
in excess of the ground-state energy is associated purely with 
squeezing of the wave packet and not with the centroid.
In this case, $\Vmax\langle\cos^2(\kt X)\rangle$
oscillates between the values 
$(\langle E_\mathrm{eff}\rangle
\pm\sqrt{\langle E_\mathrm{eff}\rangle^2-E_0^2})/2$, giving the
cooling rate of
\begin{equation}
  \Delta\langle {E}_\mathrm{eff}\rangle_\mathrm{fb}
    = -8\varepsilon\sqrt{E_\mathrm{eff}^2-E_0^2}
   \label{coolingratevar}
\end{equation}
per unit time. Then the condition that this cooling rate balances the 
heating rate in Eq.~(\ref{measurementheating}) implies
\begin{equation}
  \langle {E}_\mathrm{eff}\rangle_\mathrm{ss}
    = \frac{E_0}{\sqrt{1-\beta^2}}
   \label{E0ssvar}
\end{equation}
for the steady-state energy.  But in the regime where these theories
are valid (i.e., $\langle {E}_\mathrm{eff}\rangle_\mathrm{ss}-E_0\ll 
E_0$ or
$\beta \ll 1$),
the centroid result of Eq.~(\ref{E0ss}) is always larger than the
squeezing result of Eq.~(\ref{E0ssvar}).  Thus, we expect that
the centroid limit [Eq.~(\ref{E0ss})] is the more appropriate cooling limit.

Eq.~(\ref{E0ss}) predicts that the atomic motion will cool essentially
to the ground state so long as $\beta \alt 1/2$.
Thus, as various parameters are changed, there are ``border'' values
that divide the parameter space according to whether or not
the atomic motion is cooled to the ground state.  For example,
the atom will cool to nearly the ground state so long as
the ``bang amplitude'' $\varepsilon$ is larger than the border value
\begin{equation}
  \varepsilon_\mathrm{b} = \Gamma\kt^4.
\end{equation}
Otherwise, the steady-state energy should be substantially higher,
and the atomic motion may not even cool at all.  However, this simple
theory is only valid at small energies, so such behavior should not
necessarily be predicted well by this theory.  Additionally, this 
theory does not account for how well the Gaussian estimator 
tracks the true atomic state, and thus we would expect the actual
steady-state temperatures to be higher than predicted by 
Eq.~(\ref{E0ss}).

One final modification to this cooling theory is necessary, since as
we discuss below, parity considerations show that half of the atoms
can cool at best to the first excited band of the optical potential.
To take this effect into account, we simply modify Eq.~(\ref{E0ss})
by replacing $E_0$ by the average of the ground and first excited 
band energies, $(E_0+E_1)/2$.  But in the harmonic approximation,
$E_0 = \pi$ and $E_1 = 3\pi$, so the cooling limit becomes
\begin{equation}
  \langle {E}_\mathrm{eff}\rangle_\mathrm{ss}
    = \frac{2\pi}{1-\beta},
   \label{E01ss}
\end{equation}
This modification is simple because the centroid argument above only
assumed that the atomic state is near the band with the lowest 
achievable energy.  The odd-parity atoms have the first excited state
as their lowest achievable state, and the above argument applies just
as well to states near this effective ground state.
Thus, the ensemble average amounts to simply averaging over the two
possible steady-state energies.

\section{Simulations: Adiabatic Approximation}\label{sim1}

\subsection{Stochastic Schr\"odinger Equation}

Simulating the simplified dynamics in the adiabatic approximation according to
the SME in Eq.~(\ref{adiabaticSME}) is much less numerically intensive than 
simulating the full SME given in Eq.~(\ref{scSME}). Because of this, we will 
perform an extensive analysis of the behavior of the feedback control in this 
section using simulations in the adiabatic approximation. We will 
then verify that these simulations indeed provide a good description of the 
dynamics by performing a limited number of simulations of the full SME in the 
following section.

To perform our simulations, we first note that we may view any SME as being 
generated by averaging a stochastic Schr\"{o}dinger equation (SSE), where the 
average is taken over all the signals that the observer fails to 
measure~\cite{Barchielli93, Wiseman94b}. The unobserved signals in our case are 
the spontaneous emission from the atom and the part of the cavity output that is not 
measured due to photodetector inefficiency ($\eta<1$). 
We therefore replace nature with a formal omniscient observer, 
unknown to the observer who will effect control. 
The omniscient observer measures everything that the control observer does not, and also
has access to the control observer's measurement record. From
the omniscient observer's point of view, 
the evolution of the system is described by an SSE.
We can simulate the system 
by integrating this SSE, and even though it does not provide us with 
the control observer's state of knowledge, which would require
averaging over all possible measurement results of the omniscient
observer, it does generate instances of the control
observer's appropriate measurement record. 
The control observer can then use this measurement record to obtain an 
estimate of the state of the system by integrating the Gaussian 
estimator derived above~\cite{Doherty00}. Since the ``correct''
state estimate, given by integrating 
the full SME, is not required,
we can obtain a full simulation of the feedback control 
process by integrating the SSE, which requires far fewer resources 
than integrating the SME.

The normalized Schr\"odinger equation corresponding to the 
master equation (\ref{adiabaticSME}) that we integrate numerically
is
\begin{equation}
  \setlength{\arraycolsep}{0ex}
  \renewcommand{\arraystretch}{2.1}
  \begin{array}{rcl}
    d\psiket &{}={}& \displaystyle -iH\psiket \,dt
                 -\frac{\Gamma}{4}\cos^2(2\kt X)\psiket\,dt\\&&
                 \displaystyle
                 +\frac{\Gamma}{2}\langle\cos(2\kt X)\rangle 
                    \cos(2\kt X)\psiket\,dt\\&&
                 \displaystyle
                 +\frac{\Gamma}{2}\langle\cos(2\kt X)\rangle^2 \psiket\,dt \\&&
                 \displaystyle
                 +\sqrt{\frac{\Gamma}{2}} \left[
                   \langle\cos(2\kt X)\rangle - \cos(2\kt 
                   X)\right]\psiket\,dW,
  \end{array}
  \label{adiabaticSSE}
\end{equation}
which generates trajectories with the same measure as the SME.
Since we are integrating the SSE in lieu of the SME, there are a few
subtleties related to the measurement noise to be accounted for.
In particular, the Wiener increment $dW$ in the SSE (\ref{adiabaticSSE})
is not equivalent to the Wiener increment in expression (\ref{mradiab})
for the measurement record
if $\eta<1$, because the measured noise does not fully represent the
``true'' noise.  Rather, the quantity $\sqrt{\eta}\,dW$ in the 
photocurrent expression
should be replaced by a measured noise $dW_\eta$, given in terms of
the full noise of the SSE by
\begin{equation}
  dW_\eta = \eta\, dW + \sqrt{\eta(1-\eta)}\, dW_\mathrm{aux},
\end{equation}
where $dW_\mathrm{aux}$ is an auxiliary Wiener increment, so that
$dW_\eta^2 = \eta dt$, and a fraction 
\begin{equation}
  dW_\mathrm{unm} = (1-\eta)\, dW - \sqrt{\eta(1-\eta)}\, dW_\mathrm{aux}
\end{equation}
of the true noise is not measured by the control observer (i.e.,
$dW_\eta + dW_\mathrm{unm} = dW$).
The photocurrent in Eq.~(\ref{mradiab}) then becomes
\begin{equation}
   d\tilde{r}(t) = - \sqrt{8\eta^2\Gamma}
     \langle 
     \cos^2(\tilde{k}X)\rangle dt + \, dW_\eta ,
   \label{mragain}
\end{equation}
and the estimated Wiener increment is given in terms of the measured
noise as
\begin{equation}
  \setlength{\arraycolsep}{0ex}
  \begin{array}{rcl}
    \dWe &{}={}& \displaystyle
           \frac{dW_\eta}{\sqrt{\eta}} + \sqrt{2\eta\Gamma}\left[
           \exp(-2\kt^2\Vxe)\cos(2\kt\Xe)\right. \\&& 
            \hspace{40mm}{}- \displaystyle\left.
           \langle\cos(2\kt X)\rangle\right] dt,
  \end{array}
  \label{dWefromdWmod}
\end{equation}
which still reflects the difference between the actual and estimated
quantum states in the same way as before.

\subsection{Details of the Simulations}\label{sec:details}

For the simulations, we choose a ``canonical'' set 
of experimentally realistic parameters that will put us in the regime 
where adiabatic elimination of the cavity and internal atomic states
should be valid. 
Using the Cesium D$_2$ line as the 
atomic transition, we have 
$m=2.21\times 10^{-25}\;\mathrm{kg}$ and
$\omega_0/2\pi=351.7\;\mathrm{THz}$; we assume that the cavity
subtends a small solid angle and thus that the spontaneous
emission rate is given approximately by the free-space value of
$\gamma/2\pi=5.2\;\mathrm{MHz}$. 
For the cavity parameters, 
we choose values similar to those used in recent 
CQED experiments \cite{Mabuchi99, Hood00, Hood98},
yielding an energy decay rate of $\kappa/2\pi=40$~MHz, 
an atom-field coupling constant of $g/2\pi = 120$~MHz, 
and a mean intracavity photon number of $\alpha=1$.
However, we will consider a much larger detuning of 
$\Delta/2\pi = 4\;\mathrm{GHz}$ to the red of the atomic resonance 
than was used in the experiments, in order to work in a dispersive
regime where the adiabatic approximation should work well.
The  scaled parameters corresponding to these physical values are
$\Gamma=23.6$ and $\kt = 0.155$.  The scaled depth of the
optical potential is $V_\mathrm{max} = \pi/\kt^2 = 131$, which 
corresponds to an atomic speed of $14.7\;\mbox{cm s}^{-1}$ and
is of sufficient depth that the lowest two band energies
$E_0 = 3.12$ and $E_1 = 9.33$ are close to the corresponding values
in the harmonic approximation of $\pi$ and $3\pi$, respectively
(with
27 trapped bands in the optical potential and a ground-band width of
$1.7\times 10^{-12}$, so that tunneling transitions between wells are
highly suppressed).
Also, for the canonical parameter set we choose a feedback
amplitude $\varepsilon=0.1$ (easily accomplished experimentally)
and the idealized detection
efficiency of $\eta = 1$.  The goal of our analysis will be to
understand the dynamics of the system under these ideal conditions
and then to understand how variations in these parameters 
influence the dynamics.

The optical potential in the simulations spanned 24 wells, 
corresponding to the cavity lengths in the
Caltech experiments \cite{Hood98}.  To crudely model the
sticking behavior of an atom when colliding with a mirror,
we implemented absorbing boundary conditions, where the
absorption ramped smoothly on in the last half of each
of the boundary potential wells. We explicitly conditioned
the simulations on the fact that they were not lost, mimicking
the experimental situation where a run with a lost atom would
simply be repeated.  However, given the fact that the cavity
is relatively long compared to the optical potential period,
these details do not substantially influence the results of 
the calculation.

To understand the typical dynamics of this problem, we consider
ensemble averages.
We consider the experimental situation where
atoms dropped one at a time
have substantial motional energy after loading into the optical
potential, but only atoms that are well trapped are candidates for 
further cooling.  Thus, we assume that all atoms in the simulation 
ensemble have a centroid energy 
of $84.2$, which corresponds to a 
stationary atom at a distance of 6 away from the bottom of a well
(where the well period is $\pi/\kt = 20.3$), 
and we distribute the initial locations of the atoms uniformly 
between the bottom of the well and this maximum distance.
The atoms in the simulation begin
in a coherent state with $V_x = V_p = 1/2$
in the center well of the cavity. We do not expect that the 
steady-state results are sensitive to the details of these initial 
conditions.
Except where noted otherwise we plot averages over 128 trajectories;
in plots where we inspect the variation of ensemble averages as
a function of a parameter, each ensemble average is computed with
respect to an independent set of random-number seeds.

We obtained numerical solutions to the SSE (\ref{adiabaticSSE})
using the order 1.5 (strong), implicit, 
stochastic Runge-Kutta (SRK) algorithm
in Ref.~\cite{Kloeden92} as part of an operator-splitting
method.  To evolve the system over a large step of 
$\Delta t = 0.0005$, we first used the SRK evolver to evolve the wave
function according to the SSE without the kinetic $\pi P^2$ term
in substeps of size $\Delta t/8$ over an interval $\Delta t/2$.
Next, we applied the operator $\exp(-i\pi P^2\Delta t)$ after
a Fourier transform to momentum space.  Finally, after transforming
back, we again evolved according to the SRK algorithm by another
interval $\Delta t/2$ (in four steps) to complete the full $\Delta t$ step.
This operator splitting preserves the order 1.5 (temporal)
convergence of the SRK approximation.
The spatial grid contained 2048 points. The space and time
discretizations were found to give adequate convergence when
computed in 32-bit precision for most of the trajectories,
although a minority (typically only a few percent) displayed
some sensitivity to these numerical parameters due to their
proximity to unstable hyperbolic points in phase space.
However, for the cooling simulations, the steady-state
energies were well converged and not affected by this sensitivity,
since the atoms only saw the harmonic portions of the potential wells.

\begin{figure}[tb]
  \begin{center}
     \includegraphics[scale=0.52]{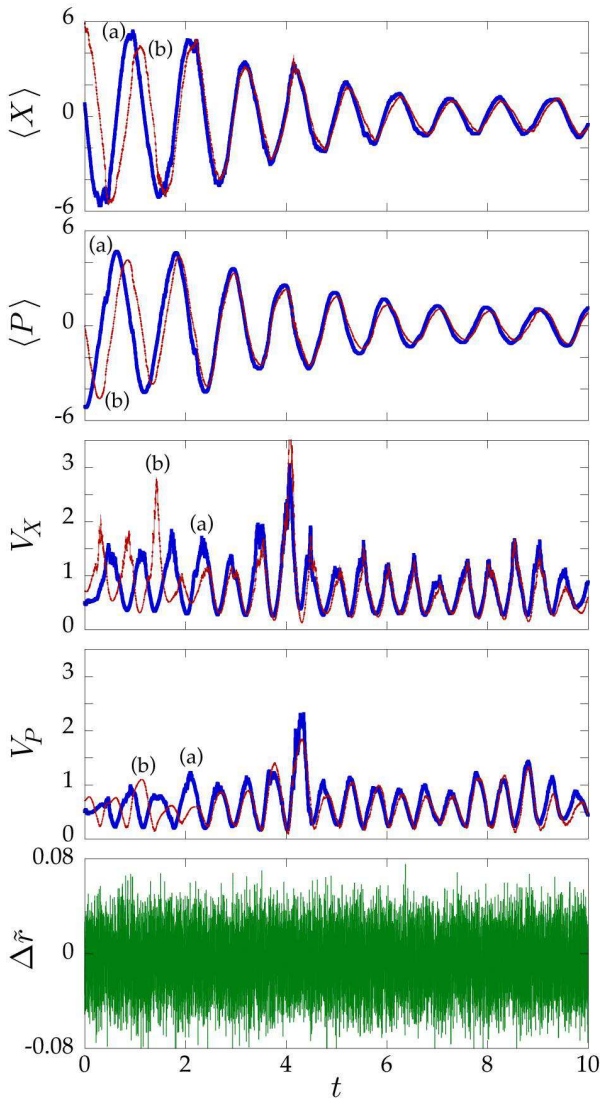}
  \end{center}
  \vspace{-5mm}
  \caption
        {(Color online) Example of locking behavior of the Gaussian estimator
         for centroids and variances.  The heavy lines (a) are the
         expectation values for the ``true'' wave packet, while the
         light lines (b) are the corresponding Gaussian estimator
         quantities.  The improved feedback cooling algorithm
         was switched on at $t=2$.  The bottom graph shows the
         corresponding photodetector signal increments $\Delta\tilde{r}$,
         plotted for the time increments of $\DtG = 0.0005$
         used by the estimator.
	\label{fig:basictracking}}
\end{figure}

The Gaussian estimator begins in each case with the impure-state
initial conditions $\Xe = 6$, $\Pe = 0$, $\Vxe = \Vpe = 1/\sqrt{2}$,
and $\Ce = 0$.  
As we argued above in Section~\ref{section:gest},
this choice of the initial estimate
is arbitrary but sufficient for our purposes, and we will see
in the next section that the cooling performance is insensitive
to any ``reasonable'' initial estimate.
The estimator is updated  with a time step of $\DtG=\Delta t$
using the simple Euler method \cite{Kloeden92}
(both to keep the calculations simple and 
because, as we discussed in Section~\ref{section:gest},
$\dWe$ is not the usual Wiener increment and thus violates assumptions
used in constructing higher-order integrators).
Some extra measures are needed to guard against instabilities
in the estimator evolution, which are due
to two possible reasons.  First, the
low-order numerical integration of the stochastic estimator equations 
with a relatively large time step tends to be unstable, especially
when the occasional large fluctuation occurs.  Second,
while Gaussian approximations to Hamiltonian evolution (or even 
evolution under an unconditioned Lindblad master equation) 
can be thought of as resulting from a variational principle, and
hence inherently stable, the Gaussian approximation to the
conditioned SME cannot, and thus does not inherit these strong 
stability properties.
We explicitly detect the four unphysical conditions: negative
position variance ($\Vxe<0$), negative momentum variance ($\Vpe<0$),
phase-space area below the Heisenberg limit 
($\Ae := [\Vxe\Vpe-\Ce^2]^{1/2} < 1/2$), 
and large position variance ($\Vxe \gg 1$, indicating the atom is not 
localized within the cavity).
If any of these conditions are detected, the variances are reset
according to the initial values $\Vxe = \Vpe = 1/\sqrt{2}$, $\Ce = 
0$, while the means $\Xe$ and $\Pe$ are not modified; the measurement
information will quickly restore the correct estimator values.
For a typical
cooling simulation with the canonical parameters, the estimator
must be reset in about 95\% of the trajectories, with trajectories
on average requiring 3 resets due to the area condition (the other
conditions producing negligible numbers of resets in comparison).
Also, to help alleviate the number of resets caused by this condition,
we instead implement the condition that resets only occur when $\Ae  < 1/4$.
This situation may seem curious, as the Gaussian area evolves
according to
\begin{equation}
  \setlength{\arraycolsep}{0ex}
  \renewcommand{\arraystretch}{1.4}
  \begin{array}{rcl}
    d(\Ae^2) &{}={}& \Gamma\kt^2\Vxe[1-\exp(-8\kt^2\Vxe)\cos(4\kt\Xe)] dt\\
    && -8\eta\Gamma\kt^2\Vxe\Ae\exp(-4\kt^2\Vxe)\sin^2(2\kt\Xe) dt\\
    && -8\eta\Gamma\kt^4\Vxe^2[1-4\kt^2\Vxe\exp(-2\kt^2\Vxe)]\\
       &&\hspace{2mm}\times \exp(-2\kt^2\Vxe)
       \cos^2(2\kt\Xe) dt\\
    &&  -\sqrt{2\eta\Gamma}\kt^2\Vxe[1-4(\kt^2\Vxe+\Ae)\exp(-2\kt^2\Vxe)]\\
       &&\hspace{2mm}  \times     \cos^2(2\kt\Xe) \dWe  ,
  \end{array}
\end{equation}
whence it follows after some examination that $\Ae \ge 1/2$ for all 
time if this is true initially.
Thus, these resets are numerical artifacts of evolving the estimator using
a finite time step and a low-order numerical method;
in the simulations, reducing the time step $\DtG$ by a factor of 10
reduces the average number of resets due to the area condition to 1.1
(with 0.8 resets on average due to the $\Vxe<0$ condition).  However, we
continue to use the larger step size, 
which is more reasonable to 
implement in practice given technological constraints on the speed of 
calculations for the forseeable future.
For somewhat smaller measurement efficiencies,
the estimator requires fewer resets (e.g., about 12\% of the 
trajectories for $\eta = 0.8$, and only 1.7 resets on average per 
trajectory), since the steady-state area is 
larger (corresponding to a state with lower purity, as 
$\mathrm{Tr}[\rho_\mathrm{e}^2] = 1/(2\Ae)$ \cite{Zurek93}, where
$\rho_\mathrm{e}$ is the estimated density operator).

Because the
estimator locks on quickly with the canonical parameters, the feedback
cooling algorithm is turned on at time $t=2$.  For the improved 
cooling algorithm, we find that fitting to the last 300 observations
(spaced apart in time by $\DtG$) is optimal.

\subsection{Estimation and Cooling Dynamics}
\label{sec:estimation-cooling-dynamics}

\begin{figure}[tb]
  \begin{center}
     \includegraphics[scale=0.41]{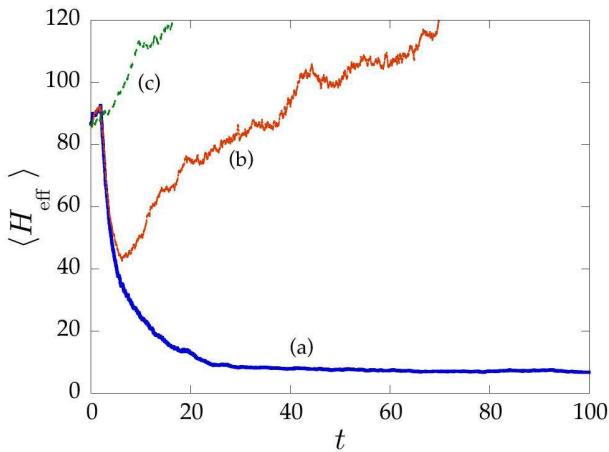}
  \end{center}
  \vspace{-5mm}
  \caption
        {(Color online) Evolution of the mean energy 
         $\langle H_\mathrm{eff}\rangle$ (relative to the
         minimum potential energy), comparing the performance
         of different cooling algorithms.
         Heavy solid line (a): improved cooling algorithm.
         Solid line (b): centroid-only cooling algorithm.
         Dashed line (c): no cooling algorithm.  Each line is an
         average of 128 trajectories; parameters are for the
         canonical set.
	\label{fig:basiccooling}}
\end{figure}

Now we will evaluate the performance of the two main components of the
quantum feedback cooling system, the estimator and the cooling
algorithm. Henceforth, we will only study the performance of the 
improved cooling algorithm, except where noted otherwise.
Fig.~{\ref{fig:basictracking} shows a typical trajectory
undergoing cooling}
and compares some of the ``true'' moments computed from the atomic
wave function (i.e., those computed with respect to the result of 
integrating the SME of the omniscient observer)
with the moments from the estimator.  The centroids 
and variances of the estimator lock on by the time that the cooling
algorithm is switched on ($t=2$), and the estimator provides a
reasonably faithful representation of the atomic dynamics.
It is especially important that the variances track the atomic
wave packet accurately in order to make the improved cooling
work effectively.  
Fig.~\ref{fig:basictracking} also shows the measurement record
increments 
\begin{equation}
  \Delta\tilde{r}(t) := \int_t^{t+\DtG} d\tilde{r}(t)
\end{equation}
corresponding to the time increments $\DtG = 0.0005$ used in
the Gaussian-estimator integration.  The measurement record
appears to be noise-dominated, but nevertheless contains
information about the quantum state in the time-dependent offset
level of the noise.
The action of the estimator is such that it ``demultiplexes'' the
information about the motions of the centroids and variances, even though they
are encoded in the same frequency range in the photocurrent signal.
One can get a sense for how this works from the form of Eqs.~(\ref{gest}).
For example, the measurement (\dWe) terms in the $d\Xe$ and $d\Vxe$
equations are weighted by $\sin(2\kt\Xe)$ and $\cos(2\kt\Xe)$, 
respectively.  Thus, the incoming measurement information 
influences $\Xe$ or $\Vxe$, depending on the state of the estimator:
if the estimator is centered in a well, variations in 
$\langle\cos^2\kt X\rangle$ (i.e., the photocurrent) are presumed
to be caused by variations in $V_x$, while such variations are
attributed to $\langle X\rangle$ if the estimator is centered on the
side of a well.
The estimator can also gain information from the 
absolute value of the measurement record---as we discussed
in Section~\ref{section:limits}, energy in only the variance 
degree of freedom can produce transient values of $\Vmax\langle[1-
\cos^2 (\kt X)]\rangle$ below $E_0/2$, whereas energy
in only the centroid cannot.
This subtle unraveling of the measurement information into the
estimated state is automatic in the measurement formulation we
have used here.

\begin{figure}[tb]
  \begin{center}
     \includegraphics[scale=0.41]{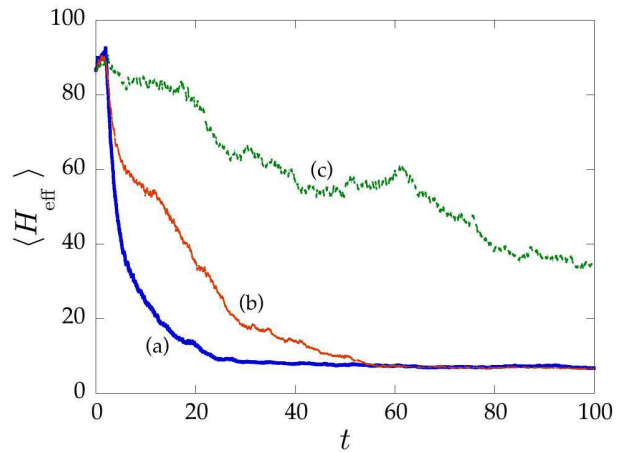}
  \end{center}
  \vspace{-5mm}
  \caption
        {(Color online) Evolution of the mean energy 
         $\langle H_\mathrm{eff}\rangle$, comparing the 
         performance of the improved cooling algorithm with
         different initial (Gaussian) state estimates.
         Heavy solid line (a): estimate matches the canonical parameter
         set.
         Solid line (b): initial estimate is much more impure and 
         substantially overlaps $(x,p)=(0,0)$.
         Dashed line (c): same as solid line, but centered on $(x,p)=(0,0)$.
	\label{fig:initcondcheck}}
\end{figure}

Fig.~\ref{fig:basiccooling} compares the ensemble-averaged 
evolution of the energy 
$\langle E_\mathrm{eff}\rangle$ in the case where the atoms are
not cooled to the cases where the atoms are cooled by the centroid-only
and improved cooling algorithms.  When the atoms are not actively
cooled, the energy simply increases linearly in time.
The initial heating rate of $\partial_t\langle 
E_\mathrm{eff}\rangle=1.78$
predicted by Eq.~(\ref{uniformheating}) for measurement-backaction
is substantially smaller than the initial simulated heating rate of
2.1(1).  We can understand this discrepancy, since the initial
conditions in the simulation are uniformly distributed in phase
space along a constant-energy surface, weighting more
heavily the gradients of the
potential where heating is maximized; Eq.~(\ref{uniformheating})
underestimates the heating rate by assuming a uniform distribution in
configuration space.

The simple, centroid-based cooling algorithm cools the atoms rapidly
for short times, but around $t=10$, the cooling stops and 
the atoms begin to heat back up, apparently without bound.
Clearly, we expect some heating from the squeezing effects 
discussed above. Furthermore, we expect that when the centroid
cools and the variance becomes much larger than the mean-square
amplitude of centroid motion, the centroid evolution will be dominated
by measurement (projection) noise, and thus the feedback signal will 
also be determined mostly by random noise from the measurement.  
In this state, the cooling algorithm is ineffective and leads to 
heating.
This process runs away, since there
is no mechanism to cool the energy associated with the variances
(squeezing).

\begin{figure}[tb]
  \begin{center}
     \includegraphics[scale=0.41]{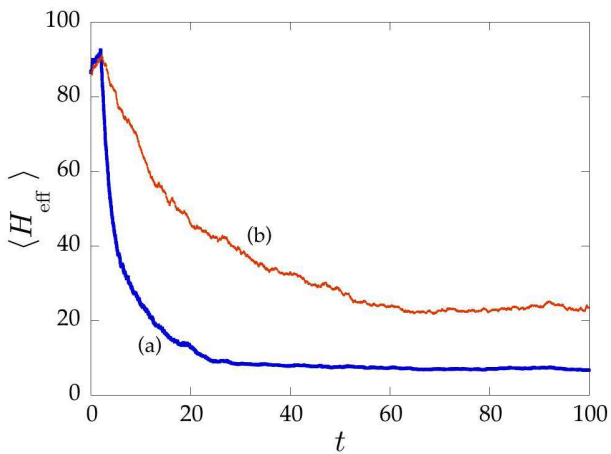}
  \end{center}
  \vspace{-5mm}
  \caption
        {(Color online) Evolution of the mean  energy 
         $\langle H_\mathrm{eff}\rangle$, comparing the
         performance of the improved cooling algorithm
         (heavy solid line, a) with that of cooling based
         directly on the homodyne signal (solid line, b).
         Heavy solid line: improved cooling algorithm.
         Each line is an
         average of 128 trajectories; parameters are for the
         canonical set.
	\label{fig:directcooling}}
\end{figure}

On the other hand, the improved cooling algorithm also rapidly
cools the atoms and the atomic energies settle to a well-defined 
steady state,
which already demonstrates the potential utility of 
this cooling algorithm for long-term storage of the atoms.
For the moment, we will defer the question of how closely the
atoms are cooling to the lowest energy band of the optical
potential until we take a closer look at the atomic dynamics 
in the next section.

The cooling behavior that we see is also insensitive to the
exact choice of the initial (Gaussian) state estimate, so long
as we have made a reasonable choice.  To illustrate this statement,
Fig.~\ref{fig:initcondcheck} shows the ensemble-averaged cooling
dynamics for two initial state estimates in addition to the 
canonical estimate that we use for the rest of the simulations.
In the first modified initial estimate, we have selected
$\Xe = 0.1$, $\Pe = 0$, $\Vxe = 1$, $\Vpe = 1$, and $\Ce = 0$
for a much larger uncertainty product (lower purity $\mathrm{Tr}[\rho^2]$)
and a substantial overlap with the other side of the potential well.
We see that although the short-time cooling performance is slightly 
worse, the long-time cooling performance is the same as for the
original initial estimator.  This implies that the tracking behavior
is initially worse for this initial estimator, but at sufficiently
long times the estimator ``forgets'' its initial state and does a 
better job of tracking.  The other initial estimator is the same 
as the previous case, but with $\Xe=0$, so that the estimator is
centered on the origin in phase space.  Examination of the 
estimator equations (\ref{gest}) reveals that in this case the
centroids $\Xe$ and $\Pe$ will remain zero for all time, effectively
``freezing out'' the estimator's centroid degree of freedom.
The cooling behavior here is much worse than in the other two cases,
demonstrating both that the estimator's variance degree of freedom is 
insufficient to mimic the dynamics of the true state and that it is
important to exercise some caution when selecting an initial state
estimate.  

Finally, we return to the question raised in 
Section~\ref{section:alg} regarding the necessity of using
the Gaussian estimator at all, in light of the fact that the
measurement record itself provides a direct, albeit very noisy,
measurement of the quantity $\langle \cos^2\kt X\rangle$ used
to determine the control switching times in the improved 
cooling algorithm.  Fig.~\ref{fig:directcooling}
address this question by comparing the energy evolution under the
improved cooling algorithm using the Gaussian estimator with 
the same algorithm but based directly on the measurement record.
In both cases, the quadratic curve was fitted on-the-fly
to the last 300 points in the measurement record to reduce noise
effects.  
We found this number of points to be optimal for the canonical
parameter set, and this optimal time interval over which the data
are fit is tied directly to the time scales of the atomic motion.
Although the direct-cooling case cools rapidly and settles to a lower
temperature, it is clear that the cooling performance
is greatly improved by the Gaussian estimator.  The estimator
acts as a nearly optimal noise filter, 
efficiently extracting the relevant information from the 
noisy measurement record.

\subsection{Parity Dynamics}

We now consider the evolution of the
parity of the conditioned atomic state due to the measurement.
For this discussion, we will focus on the expectation values 
of the usual parity operator
\prty\ (where $\prty\psi(x) = \psi(-x)$). For numerical purposes,
though, we will instead use the  \textit{reduced} parity operator,
$\prty':=\prtyr$, where the reduction operator \reduceop\ is defined by
\begin{equation}
  \reduceop\psi(x) := \sum_{j=-\infty}^\infty \psi\left(x-\frac{j\pi}{\kt}\right),\; 
    x\in\left[-\frac{\pi}{2\kt},\frac{\pi}{2\kt}\right),
\end{equation}
which allows us to detect the 
parity of the atomic state with respect to $any$ potential well in the
cavity, rather than just the center well.  
We will simply use the notation ``\prty'' for both operators
in the discussion, though, as the discussion holds for either operator.
The initial conditions for the simulations, which correspond to
Gaussian wave packets displaced in phase space from the origin
in phase space, represent equal mixtures of odd and even parity
($\langle \prty\rangle = 0$) to a good approximation.
As time evolves, however, the parity of the conditioned 
atomic state evolves (in the adiabatic approximation) according to 
\begin{equation}
  d\langle\prty\rangle = -\sqrt{8\eta\Gamma}
    \left[\langle\prty\cos(2\kt X)\rangle - \langle\prty\rangle
      \langle\cos(2\kt X)\rangle\right]\,dW .
  \label{dparity}
\end{equation}
This effect is due only to the nonlinearity of the measurement
term in the master equation (\ref{adiabaticSME}); the parity
under unconditioned evolution is invariant.

\begin{figure}[tb]
  \begin{center}
     \includegraphics[scale=0.41]{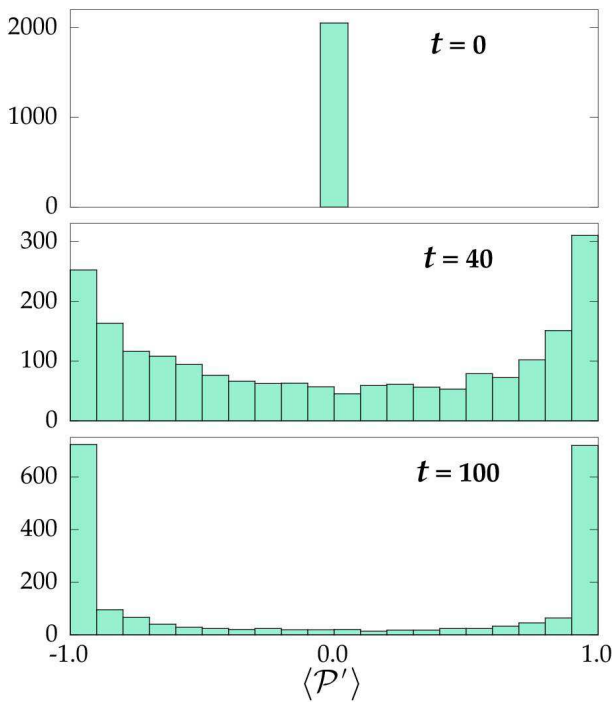}
  \end{center}
  \vspace{-5mm}
  \caption
        {(Color online) Histograms of the reduced parity $\langle\prty'\rangle$
         for an ensemble of 2048 trajectories undergoing 
         cooling, showing how parities evolve towards 
         the extreme states of pure parity at late times.
	\label{fig:parity}}
\end{figure}

The reduced parity evolution for a simulated ensemble of trajectories
undergoing measurement and cooling is shown in Fig.~\ref{fig:parity}. 
It is clear that $\langle\prty\rangle$ evolves towards the extreme
values $\pm 1$ of parity for the simulated trajectories.  
This ``parity purification'' behavior is not immediately clear
from the evolution equation (\ref{dparity}), which states that
$\langle\prty\rangle$ evolves according to a diffusion process. 
The key point is that this diffusion process is nonstationary,
since the diffusion rate depends on the parity, and indeed vanishes
for states with pure parity.  To see the final steady state directly,
we consider the projectors $\prty_+$ and $\prty_-$ for even and
odd parity, respectively, defined by
\begin{equation}
  \prty_\pm := \frac{1\pm\prty}{2}.
\end{equation}
Then the product of the expectation values of these projectors
evolves according to
\begin{equation}
    d\left[\langle\prty_+\rangle\langle\prty_-\rangle\right] = 
      -8\eta\Gamma
      \left(\langle\prty_+\rangle\langle\prty_-\rangle\wp\right)^2 
       dt,
  \label{dampparity}
\end{equation}
where the parity difference $\wp$ is given for states with
$\langle\prty_\pm\rangle\neq 0$ by
\begin{equation}
  \wp := \frac{\langle\prty_+\cos^2\kt X\rangle}{\langle\prty_+\rangle}
     - \frac{\langle\prty_-\cos^2\kt X\rangle}{\langle\prty_-\rangle}.
\end{equation}
The magnitude of $\wp$
represents how well the measurement of $\cos^2\kt X$ can
resolve the parity of the atomic state, and it is nonzero for
generic states.
Thus, the quantity $\langle\prty_+\rangle\langle\prty_-\rangle$
damps monotonically to zero, which is only possible if
the parity of the state becomes either pure even or pure odd.  Given
the initial conditions in the simulations, and the fact that the
diffusion process in Eq.~(\ref{dparity}) is unbiased, we can expect
that half of the trajectories will become even and half will become 
odd.  As we mentioned above, this argument also applies to both the
standard and reduced parity operators.
However, the damping rate in Eq.~(\ref{dampparity}) is greatly reduced
for the standard parity operator when considering 
states localized in potential wells not centered on $X=0$.

\begin{figure}[tb]
  \begin{center}
     \includegraphics[scale=0.41]{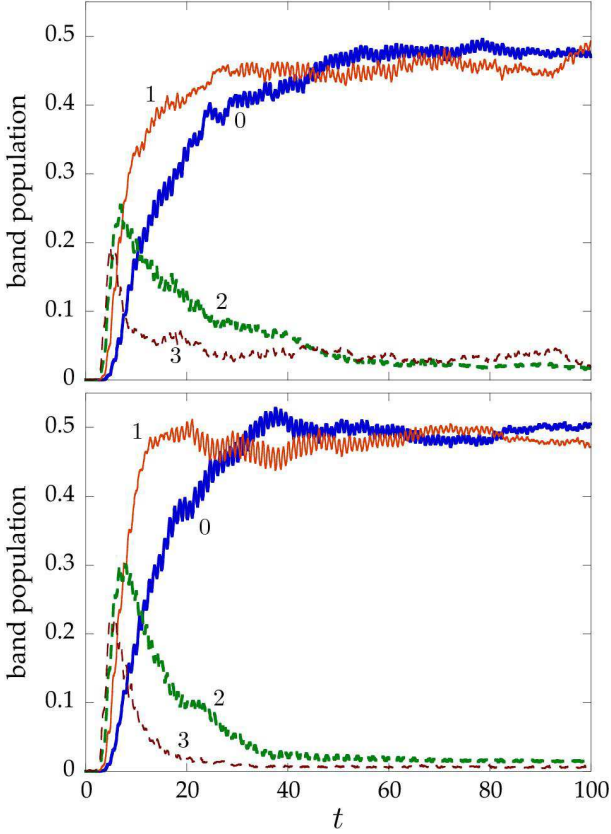}
  \end{center}
  \vspace{-5mm}
  \caption
        {(Color online) Evolution of the band populations for the lowest
         two (0 and 1) and next two (2 and 3) energy bands
         of the optical lattice, averaged over 128 trajectories.
         Top: cooling based on the Gaussian estimator. Bottom:
         cooling based on perfect knowledge of the actual
         wave function. 
	\label{fig:bandpops}}
\end{figure}

This purification effect obviously has consequences for cooling the
atomic motion.  Even if the cooling algorithm can, in principle, cool
the atomic motion faster than it is heated by the measurement, the 
cooling algorithm cannot affect the parity of the atomic state, and 
thus odd-parity states will not be cooled to the ground state of the 
optical potential.
At best, then, the cooling algorithm can cool the atom to the 
ground energy band with probability $1/2$ and the first excited
energy band with probability $1/2$.  
This cooling behavior is illustrated in Fig.~\ref{fig:bandpops},
where the evolutions of the band populations are plotted for the 
two
cases where cooling is based on the Gaussian estimator and 
directly on the true atomic wave function using the improved 
algorithm in both cases.  We can clearly see that in steady state
the lowest and first excited bands are about equally
populated, with these bands accounting for 94\% 
of the population
in the Gaussian-estimator case and 98\% of the population in the
perfect-knowledge case.  Thus, this parity effect is the 
dominant limit to the atomic motional energy for the ensemble.
However, this situation is essentially
as good as cooling fully to the ground band, because the measurement
can distinguish between the two possible outcomes (i.e., by resolving
the two different possible steady-state potential energies).  Experimentally,
if the odd-parity outcome is detected, the atom can be transferred to the
ground band by a coherent process (e.g., resonant modulation
of the optical potential by an external field or two-photon, stimulated
Raman transitions), or the state can simply be rejected, in hopes of
obtaining the even-parity outcome on the next trial.


Finally, we comment on one odd aspect of the
intermediate-time histogram ($t=40$) in Fig.~\ref{fig:parity}, which
shows a marked asymmetry, even though 
the \textit{median} value of $\langle\prty\rangle$ is zero.  
Because we can write
Eq.~(\ref{dparity}) in the form
\begin{equation}
  d\langle\prty\rangle = -\sqrt{8\eta\Gamma}
    \langle\prty_+\rangle\langle\prty_-\rangle\wp\,dW ,
  \label{dparityagain}
\end{equation}
all of the asymmetry is due to the $\wp$ factor.  Since this 
diffusion process mimics a quantum nondemolition measurement of the 
parity, we can interpret $\wp$ as an effective measurement strength
that is larger for even-parity states than for odd states.  This 
asymmetry is reasonable since the measurement and cooling processes produce
even-parity states that are more localized than the odd ones.  Thus, 
the even-parity states, being tightly localized at the field antinodes,
produce more extreme values of the measurement record and are hence 
more easily distinguished from mixed-parity states.
Again, however, this effectively asymmetric parity-measurement process
does not affect the long-time outcome that either parity will be 
selected with equal probability.



\subsection{Tests of Cooling-Limit Theories}

\begin{figure}[tb]
  \begin{center}
     \includegraphics[scale=0.41]{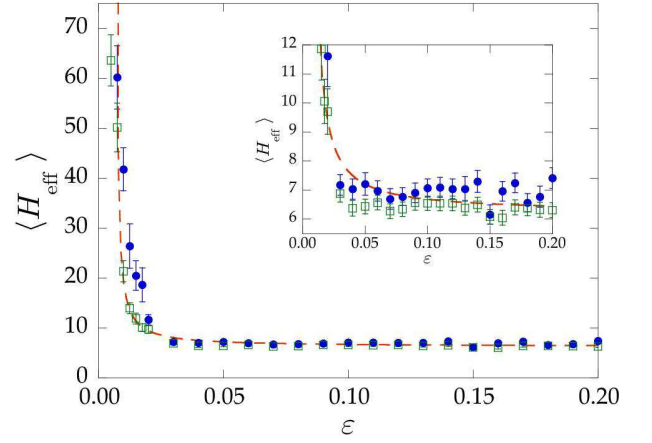}
  \end{center}
  \vspace{-5mm}
  \caption
        {(Color online) Final energies, measured over the interval from $t=90$
         to $100$, as the switching amplitude $\varepsilon$
         varies; other parameters match the canonical set.
         Circles: cooling based on the Gaussian estimator.
         Squares: cooling based on perfect knowledge of the 
         actual wave function.  Dashed line: simple cooling 
         theory, Eq.~(\ref{E01ss}).  Inset: magnified view of the
         same data. Error bars reflect
         standard errors from averages over 128 trajectories.
	\label{fig:epsilonsweep}}
\end{figure}

\begin{figure}[tb]
  \begin{center}
     \includegraphics[scale=0.41]{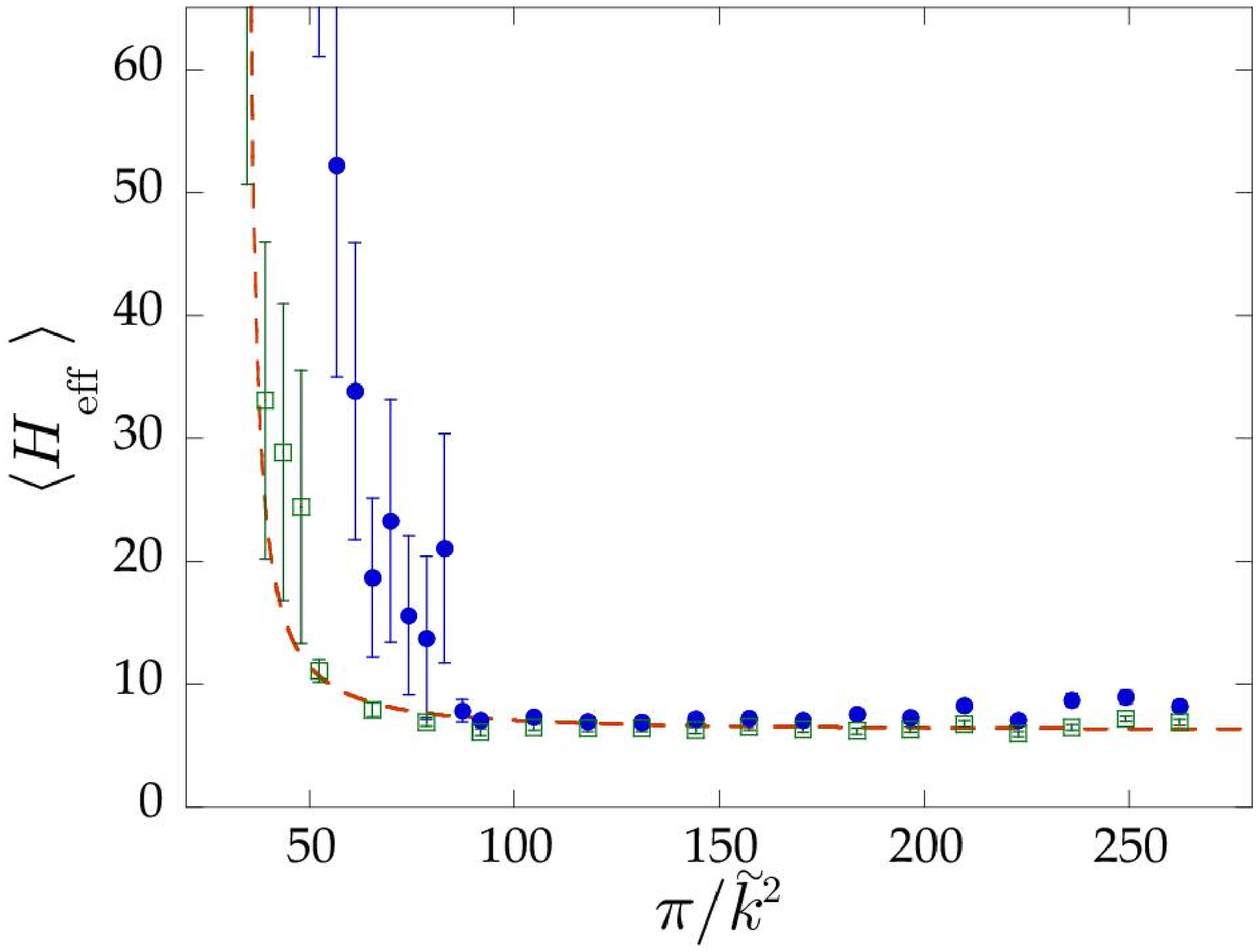}
  \end{center}
  \vspace{-5mm}
  \caption
        {(Color online) Final energies, measured over the interval from $t=90$
         to $100$, as the unmodulated optical potential depth 
         $\Vmax = \pi/\kt^2$
         varies; other parameters match the canonical set.
         Circles: cooling based on the Gaussian estimator.
         Squares: cooling based on perfect knowledge of the 
         actual wave function.  Dashed line: simple cooling 
         theory, Eq.~(\ref{E01ss}).  Error bars reflect
         standard errors from averages over 128 trajectories.
	\label{fig:vmaxsweep}}
\end{figure}

Now we can examine the dependence of the steady-state cooling
energies on the system parameters and test the validity of the
cooling-limit theories in Section~\ref{section:limits}.
Fig.~\ref{fig:epsilonsweep} shows the dependence of the
simulated steady-state temperatures on the control-switching
amplitude $\varepsilon$ in both the Gaussian-estimator
and perfect-knowledge cases.  The simple cooling prediction of 
Eq.~(\ref{E01ss}) is also shown here for comparison.  
The agreement between the theory and the perfect-knowledge
simulations is much better than between the theory and
the Gaussian-estimator simulations, which is sensible because
the simple theory does not attempt to account for the
imperfections in how the estimator tracks the true state.
While the (perfect-knowledge) steady-state energies consistently
lie slightly below the theoretical prediction, the agreement
is overall quite good, and the theory correctly predicts 
a sharp transition from efficient to inefficient cooling
near the transition border
$\varepsilon_\mathrm{b} = 0.014$
(corresponding to the point $\beta\sim 1/2$ where the predicted
steady-state energy is twice the smallest predicted value).
Simply put, when the control transitions of the optical
potential are sufficiently weak, the cooling algorithm does
not cool the atoms at a rate sufficient to counteract the measurement
heating (the system is no longer ``controllable''), and so the atoms do not equilibrate at a
very low temperature.
The cooling and transition behaviors are very similar in the
Gaussian-estimator simulations, but these energies are
overall slightly higher, again probably due to the inability of the
Gaussian estimator to perfectly track the atomic state, even
with $\eta = 1$.

Similar behavior appears as the (unmodulated) optical potential
depth $\Vmax = \pi/\kt^2$ varies, as shown in Fig.~\ref{fig:vmaxsweep}.
For sufficiently large potential depth, the temperature is essentially
constant, but if the potential depth is too weak, we again lose
controllability and the cooling algorithm fails to be effective.
There is a much more substantial difference between the 
Gaussian-estimator simulations and the simple cooling theory for
small $\Vmax$ compared
to Fig.~\ref{fig:epsilonsweep}.  This discrepancy is most likely
due to the reduced confinement of the wave packet, which produces 
stronger anharmonic effects and thus a breakdown of the Gaussian
approximation.
The simulation data are still in good agreement with the simple 
theory.

\begin{figure}[tb]
  \begin{center}
     \includegraphics[scale=0.41]{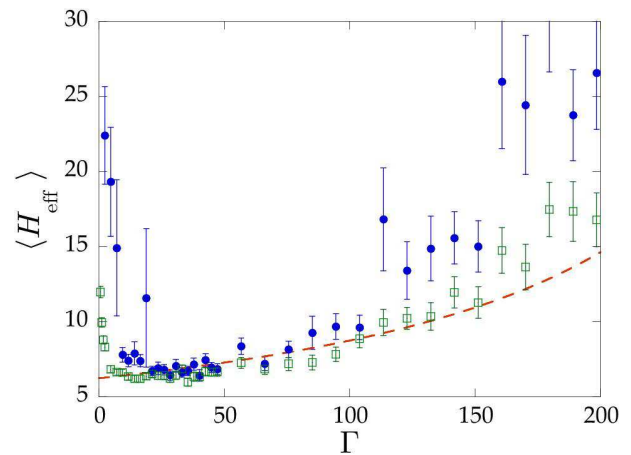}
  \end{center}
  \vspace{-5mm}
  \caption
        {(Color online) Final energies, measured over the interval from $t=90$
         to $100$, as the effective measurement strength $\Gamma$
         varies; other parameters match the canonical set.
         Circles: cooling based on the Gaussian estimator.
         Squares: cooling based on perfect knowledge of the 
         actual wave function.  Dashed line: simple cooling 
         theory, Eq.~(\ref{E01ss}). Error bars reflect
         standard errors from averages over 128 trajectories.
	\label{fig:gammasweep}}
\end{figure}

The dependence of the steady-state energy on the
measurement strength $\Gamma$ is shown in Fig.~\ref{fig:gammasweep}.
In this case, the theory again matches the perfect-knowledge
situation quite well.  The cooling is best for moderately small
values of $\Gamma$, and the simple theory predicts a transition
to an uncooled state if $\Gamma$ exceeds the border value
$\Gamma_\mathrm{b} = \varepsilon/\kt^4 = 173$, since $\Gamma$
controls the rate of heating in the system.  However, due to the
functional form of Eq.~(\ref{E01ss}), the ``transition'' here is
much more gradual than the transition observed when varying
$\varepsilon$.
The Gaussian-estimator simulations again result in values that
are consistently slightly above the perfect-knowledge energies.
The difference is very pronounced for small $\Gamma$, where we
might expect that the
estimator is not gaining sufficient information from the 
measurement and cannot track the true state well enough
to produce effective cooling (i.e., the system is no longer 
``observable'').
The perfect-knowledge energy data also show a slight upturn
for small $\Gamma$ that is not predicted by the simple theory.
This departure indicates the direct importance of the localization
produced by the measurement in the cooling process: 
a localized phase-space distribution
is responsible for larger fluctuations in $\langle\cos^2\kt X\rangle$
than a delocalized distribution with the same energy, and from the
discussion in Section~\ref{section:alg}, the
cooling rate is directly proportional to the magnitude of these
fluctuations.  Thus, a sufficiently large value of $\Gamma$ must
be chosen to maintain a localized atomic distribution and hence
good cooling.

\begin{figure}[tb]
  \begin{center}
     \includegraphics[scale=0.41]{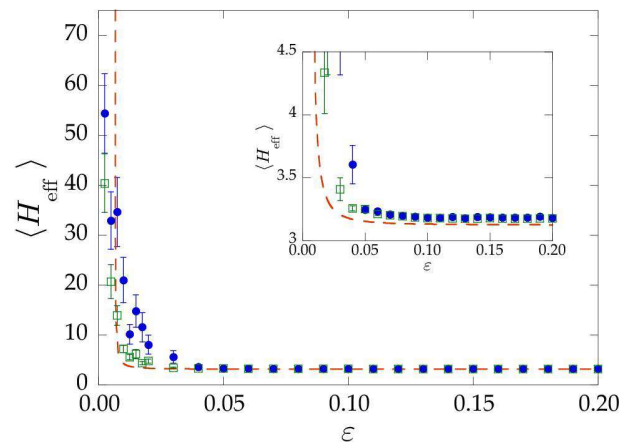}
  \end{center}
  \vspace{-5mm}
  \caption
        {(Color online) Final energies as in Fig.~\ref{fig:epsilonsweep}, but
         with initial position and momentum centroids (for both
         the atomic wave packet and Gaussian estimator) at zero.
         The dashed line now refers to the variance-dominated,
         even-parity theory of Eq.~(\ref{E0ssvar}).
	\label{fig:epsilonctrsweep}}
\end{figure}

Thus far, we have focused on testing the theory of Eq.~(\ref{E01ss}),
which is the most relevant theory for a physical implementation of
the cavity QED feedback system.
However, we can also test the variance-only theory of 
Eq.~(\ref{E0ssvar}) in the simulations, thereby giving indirect
support to the more physically relevant theory.
We can do this by changing the initial condition of the atomic
wave packets to be a coherent (Gaussian) state centered on 
one of the potential wells with zero centroid momentum.  
From the form of the Gaussian
estimator equations (\ref{E0ssvar}) we see that the wave-packet
centroid will remain fixed at zero position and momentum, and
thus all the energy will be associated with the wave-packet
variances (in the harmonic approximation).  The results of simulations
of this type, where we also modify the initial condition of the 
Gaussian estimator to have the same centroid, are plotted
in Fig.~\ref{fig:epsilonctrsweep}.  The theory of Eq.~(\ref{E0ssvar})
is also shown; note that we do not need to replace the $E_0$
in this expression with the average of $E_0$ and $E_1$, since
the initial conditions have exactly even parity, which is then
time-invariant.  The theory is again in good agreement with the
perfect-knowledge data, both in the lower temperatures for 
large $\varepsilon$ and the sharper transition and
lower transition border of
$\varepsilon_\mathrm{b} = 0.0079$ (corresponding to 
$\beta\sim\sqrt{3}/2$, again when the predicted energy is twice 
the minimum predicted value).

\begin{figure}[tb]
  \begin{center}
     \includegraphics[scale=0.41]{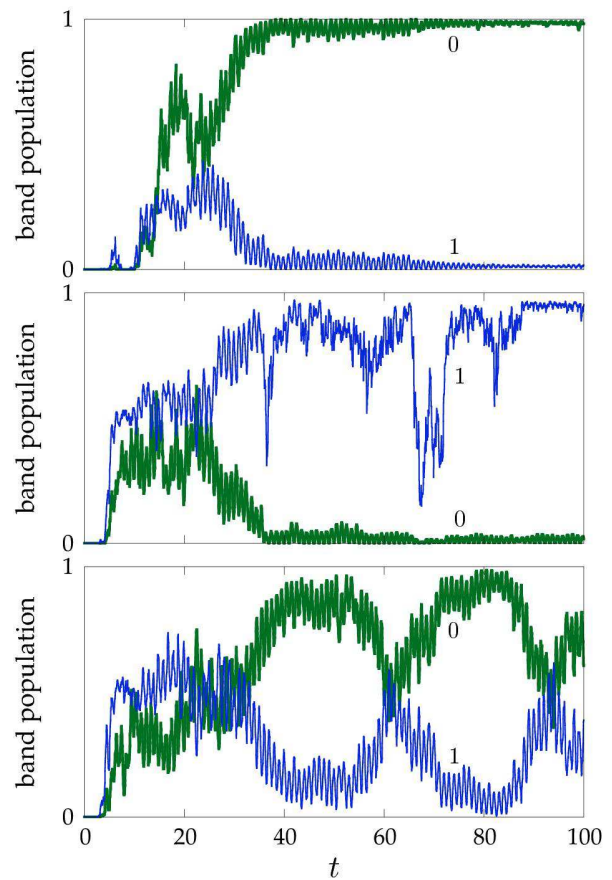}
  \end{center}
  \vspace{-5mm}
  \caption
        {(Color online) Evolution of the ground and first-excited band populations
         for three trajectories,
         illustrating typical behavior.
         Trajectories that purify to even parity (top) tend to
         settle to a quiet equilibrium in the ground state.
         Trajectories purifying to odd parity (middle) are
         characterized by dynamical equilibria, where periods
         in the first excited state are interrupted by 
         episodes of higher energy.  Trajectories that 
         take longer to purify in parity (bottom) cool
         rapidly to the lowest two states, after which 
         population is transferred between these two states
         as the parity diffuses.
         Parameters here correspond to the canonical set.
	\label{fig:bandpoptrajs}}
\end{figure}

Finally, a note about the steady-state temperatures is in order
for the cases where the Gaussian estimator is used for cooling.
The ensemble-averaged, late-time energies in 
Figs.~\ref{fig:epsilonsweep}-\ref{fig:epsilonctrsweep} 
for Gaussian-estimator cooling are slightly but
consistently higher than the corresponding perfect-knowledge
energies, suggesting that the cooling is somewhat less effective
when the Gaussian estimator is used.  However, these averages
mask complicated dynamics that are illustrated in 
Fig.~\ref{fig:bandpoptrajs}.  
Trajectories that have even parity at late times settle into
a quiet equilibrium with nearly all the population in the ground
energy band.  On the other hand, trajectories that settle to odd parity 
have a more complicated equilibrium, where the atom is mostly in the 
first excited band but the evolution is punctuated by ``bursts'' of 
heating.  This is an indication that in steady state,
the Gaussian estimator is, unsurprisingly, a much better approximation
for a wave packet close to the ground state than a wave packet close to the 
first excited state.  As seen in Fig.~\ref{fig:bandpoptrajs}, these
heating episodes correspond to population being transferred to 
higher energy states of odd parity, a process distinct from the parity 
drift which, after the initial cooling, transfers population between
the ground and first excited states.
Again, if the goal of the cooling process is to prepare an atom in
the ground state, the rate of success is even better than suggested 
by the ensemble-averaged energies, since cooling in the even-parity 
cases is essentially as good as if one had perfect knowledge of the
actual wave function.


\section{Simulations: Full Atom--Cavity Dynamics}\label{sim2}

To verify the validity of the results of this paper, it is 
important to check the adiabatic approximation that leads to the
reduced master equation (\ref{adiabaticSME}).  We thus carried
out simulations of the full, coupled atom--cavity dynamics described 
by Eq.~(\ref{scSME}).

\begin{figure}[tb]
  \begin{center}
     \includegraphics[scale=0.41]{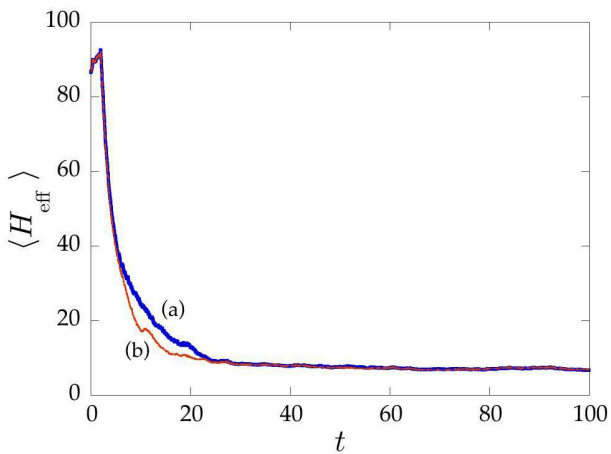}
  \end{center}
  \vspace{-5mm}
  \caption
        {(Color online) Ensemble energy evolution comparing the effect of the spatial
         grid size (cavity mirror separation) on the atomic evolution.
         Heavy solid line (a): canonical parameters (adiabatic 
         approximation) spanning 24 optical potential wells. Light solid 
         line (b):
         6 potential wells.
         Each curve represents an average over 128 trajectories.
	\label{fig:smcavtest}}
\end{figure}

\begin{figure}[tb]
  \begin{center}
     \includegraphics[scale=0.41]{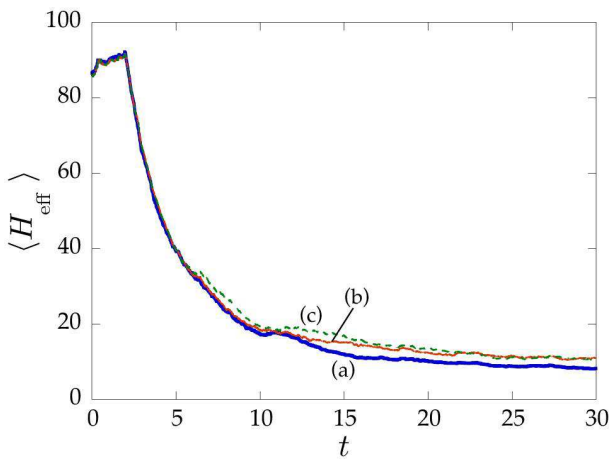}
  \end{center}
  \vspace{-5mm}
  \caption
        {(Color online) Ensemble energy evolution comparing evolution with and
         without invoking the adiabatic approximation.  
         Heavy solid line (a): adiabatic approximation. Light solid 
         line (b):
         full atom-cavity dynamics at 4 GHz detuning from atomic
         resonance. Dashed line: full atom-cavity dynamics at 2 GHz
         detuning from atomic resonance.
         Each curve represents an average over 128 trajectories.
	\label{fig:longrun}}
\end{figure}

The simulation procedure is the same as before, except that we used
the normalized Schr\"odinger equation corresponding to the 
master equation (\ref{scSME}),
\begin{equation}
  \setlength{\arraycolsep}{0ex}
  \begin{array}{rcl}
    d\psiket &{}={}& \displaystyle -iH\psiket \,dt
                 -\frac{\kapt}{2}a^\dagger a\psiket\,dt
                 +\frac{\kapt}{2}\langle a+a^\dagger\rangle a\psiket\,dt\\&&
                 \displaystyle
                 -\frac{\kapt}{8}\langle a+a^\dagger\rangle^2\psiket\,dt 
                 +\frac{\gamt}{2}\left(\langle \sigma^\dagger\sigma\rangle
                    -\sigma^\dagger\sigma\right)\psiket\,dt\\&&
                 \displaystyle
                 +\sqrt{\kapt}\, a\psiket\,dW
                 -\frac{\sqrt{\kapt}}{2}\langle a+a^\dagger\rangle\psiket\,dW
                 \\&&\displaystyle
                 +\left(\frac{\sigma e^{i\ut X}}
                      {\sqrt{\langle\sigma^\dagger\sigma\rangle}}-1\right)
                 \psiket\,dN.
  \end{array}
\end{equation}
We used the canonical parameter set described before.  
This set leads to 
scaled parameter values of $\gamt=190$ and $\kapt=1460$ in addition to
those used in the adiabatic-approximation simulations as described in 
Section~\ref{sec:details}.
The detuning
$\Delta/2\pi = 4\;\mathrm{GHz}$ from atomic resonance implies a wide 
separation of timescales in the problem.  
Thus we again evolved the system over a large step of 
$\Delta t = 0.0005$ but in much smaller substeps of $\Delta t/4096$, applying the 
kinetic evolution operator after half of the substeps.
The spatial grid contained 512 points, spanning only 6 wells to make 
the computation more tractable. The impact of the smaller grid on the 
cooling was to slightly accelerate the cooling process, as 
illustrated in Fig.~\ref{fig:smcavtest}.  This effect due to an effective 
evaporation of hot atoms, since we treat the mirrors as absorbing 
boundary condition as discussed above.
We also used 7 cavity states and 2 internal atomic states.
Due to the small time step, we found it necessary to 
use 64-bit precision for these simulations.

The simulation results are shown in Fig.~\ref{fig:longrun}, where we 
see that there is only a small difference between the cooling 
performance in the adiabatic and full simulations.  Also shown is
a simulation with a detuning $\Delta/2\pi = 2\;\mathrm{GHz}$, evolved 
with substeps of $\Delta t/2048$, to increase the nonadiabatic effects.
The parameters were scaled to compensate.  Even with this relatively 
close detuning, we see that the cooling dynamics are well 
described in the adiabatic approximation.
Other ensemble-averaged quantities corroborate this point.
For the 4 GHz detuning case, the atomic excited-state population 
$\langle \sigma^\dagger \sigma\rangle$ is
$8.4\times 10^{-3}$ (corresponding to a spontaneous emission rate of 
0.16); the adiabatic-approximation expression $\tilde{g}^2|\alpha|^2/(\tilde{\Delta}^2 + 
\tilde{\kappa}^2)$ predicts a similar value of $8.6\times 10^{-3}$ if we include a factor
of $1-\langle E_\mathrm{eff}\rangle/2\Vmax = 0.98$ to account for the
ensemble-averaged spatial distribution of the atoms.
For the 2 GHz case, the excited-state population is $3.4\times 
10^{-2}$ (spontaneous emission rate of 0.68), 
agreeing well with the adiabatic prediction of $3.5\times 10^{-2}$.
In both cases, the cavity excitation $\langle a^\dagger a\rangle = 
0.98$, in good agreement with the adiabatic value of 1.


\section{Conclusion}\label{conc}
Real-time quantum feedback control is worth investigating not only 
because of potential future applications, but because it will provide 
an undeniable litmus test of quantum measurement theory at the level 
of individual measurement records produced by measuring single 
systems. In this work we have shown that, by using a simplified 
estimation technique coupled with a relatively simple feedback algorithm, 
a single atom may be cooled close to its ground state in a realistic 
optical cavity. We have also explored the effectiveness of the feedback 
algorithm in various parameter regimes and we hope that this will help 
to guide future experiments.

\section*{Acknowledgments}
The authors would like to thank Andrew Doherty 
and Sze Tan for helpful discussions.
 This research was performed in 
part using the resources of the Advanced Computing Laboratory, 
Institutional Computing Initiative, and LDRD program
of Los Alamos National Laboratory.

\appendix

\section{Simple Feedback Cooling and Squeezing}

Here we analyze the effect of the simple centroid-only feedback cooling algorithm 
described in Section~\ref{section:alg}. 
This involves switching the optical potential to a high value when 
the atom is climbing a potential well, and switching it to a low value 
when it is falling. 
Recall that the value of the ``switched high'' field is 
$(1 + \varepsilon_1) E$ and the ``switched low'' field is 
$(1 - \varepsilon_2) E$, so that the
potential height has the corresponding values
$(1 + \varepsilon_1)^2 V$ and $(1 - \varepsilon_2)^2 V$ ($E$ and $V$ 
are the respective unmodulated values).

Since the feedback algorithm involves more than 
simply driving the system with an external force, 
to understand the action of the algorithm one must analyze its 
effect both on the means (the centroid) and the variances of the atomic position and 
momentum. The effect on the centroid is to damp the motion. In one cooling 
cycle, the energy of the centroid is multiplied by a ``cooling
factor,''
\begin{equation}
  C_{\mbox{\scriptsize fb}} = \left(\frac{1 - \varepsilon_2}{1 + 
\varepsilon_1}\right)^2, 
\end{equation}
which is valid in the harmonic approximation.  

The effect on the variances is, however, to squeeze the atomic wave function. 
To understand why this is the case, consider what happens when the 
height of the potential is changed. This changes the frequency of the 
effective harmonic oscillator governing the atomic motion, and as a 
consequence, a wave function that was not squeezed in the phase space 
of the old harmonic oscillator, is squeezed in the space of the new 
one. In particular, raising the height of the potential scales phase 
space so that  momentum is squeezed, and conversely lowering the 
potential squeezes position. To determine the rate of squeezing, it 
is useful to consider what happens in one ``cooling cycle.'' For 
example, we may define the cooling cycle to start when the atom 
crosses the bottom of the well, and end when the atom again crosses 
the bottom of the well traveling in the original direction 
(essentially one period of oscillation). Examining the effect of each 
of the changes in the potential during a cooling cycle, we find that 
the result of a single cycle is to multiply the momentum variance of 
the atomic wave function by a ``magnification factor,''
\begin{equation}
  M_\mathrm{fb} = \left(\frac{1+\varepsilon_1}{1 - 
\varepsilon_2}\right)^2,
\end{equation}
in the harmonic approximation.  This factor is precisely the 
inverse of the cooling factor $C_\mathrm{fb}$.
  
This situation is analogous to an attempt to cool a classical 
ensemble of particles using a control algorithm designed for a single
particle; although the algorithm will cool a single particle, it 
will almost certainly heat the other members of the 
ensemble. This will be an important consideration in designing an 
improved feedback algorithm. A simulation of this simple feedback 
algorithm is given in Figure~\ref{fig:basiccooling}. This shows that after 
a few cooling cycles in which the total energy is initially reduced, the 
resulting squeezing overtakes the cooling of the centroid, and the 
algorithm actually heats the atomic motion.

\section{Heating Mechanisms}\label{heating}

\noindent\textsl{\bfseries Spontaneous Emission.}
Spontaneous emission events kick the atom in both directions, and 
therefore cause momentum diffusion. The effect of this is 
to increase the energy, on average, linearly with time. 
This rate depends both upon the 
spontaneous-emission rate as well as the magnitude of the associated kicks.
One point to note is that the magnitude of each 
kick is not merely determined by the momentum of the emitted photon, 
but also by the difference between the momentum of the excited and 
ground state wave functions. 
More precisely, we can factor the atomic state vector \ket\psi\ 
just before a spontaneous-emission event
into a product of internal and motional states,
\begin{equation}
  \ket\psi = \ket{\psi_\mathrm{e}}\ket{\mathrm{e}} + 
             \ket{\psi_\mathrm{g}}\ket{\mathrm{g}},
\end{equation}
where \ket{\psi_\mathrm{\alpha}}\ are states in the center-of-mass
space of the atom, and ``e'' and ``g'' denote excited and ground atomic
energy levels, respectively. Then in an unraveling of the 
master equation (\ref{SME}) where the spontaneously emitted photons
are detected but do not give position information about the atom,
the one-dimensional atomic state just after a 
spontaneous-emission event is just the previous state with the
internal state lowered and a momentum-recoil factor,
\begin{equation}
  \ket\psi = \ket{\psi_\mathrm{e}}\ket{\mathrm{g}}e^{-i\tilde u X},
\end{equation}
where the random variable $\tilde u$ is chosen from the projected
angular distribution of the resonance fluorescence $N(\tilde u)$
referred to in the SME (\ref{scSME}).
Thus, in addition to the emission momentum recoil, the
spontaneous emission effectively converts the atomic state from the ground 
state wave function to the excited state wave function.  In the 
regime where the adiabatic approximation is valid, we have the 
relation
\begin{equation}
  \ket{\psi_\mathrm{e}} = \frac{\alpha g}{\Delta}\cos(kx)
  \ket{\psi_\mathrm{g}},
\end{equation}
and thus we can interpret this extra
process as being simply another momentum kick due to the absorption of a cavity 
photon (i.e., a superposition of a photon-recoil kick in either
direction along the cavity axis).

The actual rate of spontaneous emissions may be estimated by multiplying the 
spontaneous emission rate by the average excited state population, 
giving $\tilde{\gamma}\tilde{g}^2|\alpha|^2/(\tilde{\Delta}^2 + 
\tilde{\kappa}^2)$ for an atom localized at a field antinode. 
Calculating this for the conditions 
that we use in the numerical simulations in the body of the text, we find that the 
rate is around 0.17.  In the far-detuned regime that we study here,
the spontaneous-emission heating is a minor effect compared to 
the other heating mechanisms. 

\noindent\textsl{\bfseries Measurement Backaction.} 
The rate of heating from the back-action due to the continuous 
measurement process may be estimated by examining the effective 
master equation for the system in 
Eq.~(\ref{adiabaticSME}). 
The energy of the atomic motion in the effective 
potential, defined with respect to the minimum 
potential energy, is simply 
\begin{equation}
  E_\mathrm{eff} = \pi P^2 +
  \frac{\pi}{\kt^2}\left[1-\cos^2(\tilde{k}X)\right] .
\end{equation} 
The rate of increase of this energy, averaged over all
possible trajectories, due to the 
measurement is determined by the term in the SME
proportional to $\Gamma$, which gives
\begin{equation}
  \setlength{\arraycolsep}{0ex}
  \renewcommand{\arraystretch}{1.8}
  \begin{array}{rcl}
  \partial_t\langle {E}_\mathrm{eff}\rangle_\mathrm{meas} &{}={}& 2\Gamma 
  \mathrm{Tr}\{E_\mathrm{eff}{\cal D}[\cos^2(\kt x)]\rho\}  \\ 
          &{}={}& 8\pi\Gamma \langle \cos^2(\kt X)\sin^2(\kt X)\rangle .
  \end{array}
\end{equation}
Note that the $\Gamma$ here is actually a time-averaged quantity when
feedback is applied, which
we assume to be well approximated by the unmodulated value of $\Gamma$
in Eq.~(\ref{Gammadef}), at least to $O(\varepsilon)$.
The simplest limit for this expression is if we assume the particle 
to be randomly distributed along the potential, 
which is a reasonably good approximation to the initial heating rate:
\begin{equation}
  \partial_t\langle {E}_\mathrm{eff}\rangle_\mathrm{meas} = \pi\Gamma\kt^2 .
  \label{uniformheating}
\end{equation}
A more appropriate limit for this heating rate after cooling has taken 
place is the 
low-temperature regime, where we can make the harmonic approximation
so that the rate becomes
\begin{equation}
  \partial_t\langle {E}_\mathrm{eff}\rangle_\mathrm{meas}
    \approx 8\Gamma\kt^2 \langle \kt^2 X^2\rangle
    \approx 4\Gamma\kt^4
  \langle E_\mathrm{eff}\rangle,
  \label{measurementheating}
\end{equation}
which is valid for small $\langle E_\mathrm{eff}\rangle$.

Note that although we attribute this heating to measurement
back-action, an equivalent, measurement-independent picture for this
heating process is the cavity decay process.  Because light escapes the
cavity in discrete units at random times, there is a stochastic 
component to the cavity field intensity.  Thus, there is a 
stochastic, heating force on the atoms. 
The rate obtained by this argument must be the same as the
rate that we just obtained, since the heating rate obtained by 
averaging over all possible trajectories in a ``quantum jump''
unraveling of the master equation must be the same as in the ``quantum
diffusion'' unraveling that we use here.

\section{Effects of Detection Inefficiency and Feedback Delays}

\subsection{Detection Efficiency}

\begin{figure}[tb]
  \begin{center}
     \includegraphics[scale=0.41]{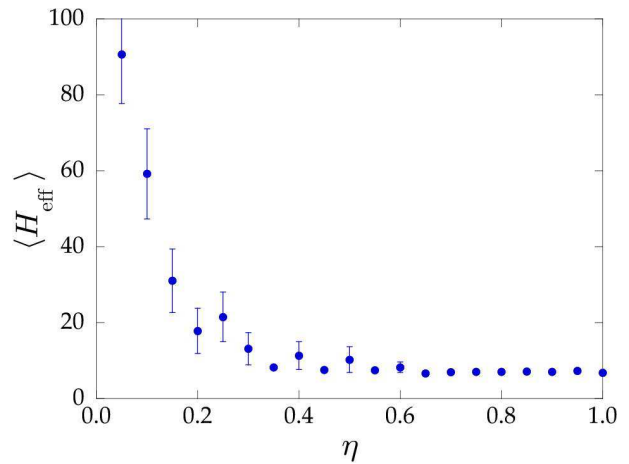}
  \end{center}
  \vspace{-5mm}
  \caption
        {(Color online) Final energies, measured over the interval from $t=90$
         to $100$, as the detection efficiency $\eta$
         varies; other parameters match the canonical set.
         Cooling is according to the improved algorithm based
         on the Gaussian estimator; error bars reflect standard
         errors from averages over 128 trajectories.
	\label{fig:etasweep}}
\end{figure}

Now we come to one of the two primary limitations in an experimental
implementation of the present system, that of limited efficiency
of the optical detectors. As this limits the information incorporated
by the estimator, it is natural to expect that the best cooling
is achieved for unit efficiency, and that the cooling simply gets 
worse for inefficient measurements.  Simulated final energies 
using the Gaussian estimator are shown in Fig.~\ref{fig:etasweep},
which for the most part bear out this expectation.  One remarkable
feature of these results is that the cooling is extremely robust
to detector inefficiency.  It is only down around 50\% detection efficiency
that a few trajectories begin to pull up the average energies,
and below about 35\% efficiency that there is a clear trend 
towards large steady-state temperatures.  This robustness is
very encouraging that an experimental demonstration of 
quantum feedback cooling will work well in spite of imperfect
detectors.

\subsection{Speed and Feedback Delay Issues}\label{section:speed}

In the simulations we have discussed so far, the Gaussian estimator
was updated every time interval of $\DtG = 0.0005$.  For the 
canonical parameter set, this update interval corresponds to about 3
ns in physical time.  This is obviously a very short time in which to
evolve the estimation equations (\ref{gest}) as well as the curve fit
outlined in Section~\ref{section:alg}.
Therefore, the first issue we will tackle is the speed with which 
one can iterate the control algorithm.

\noindent\textsl{\bfseries Curve Fit.}
Recall that, given a set of estimates $y_{\mathrm{est}, n}$ 
corresponding to 
times $t_n$, we wish to implement the curve fit of the
function $a_0+a_1 x + a_2 x^2$ to the last $q$ values of
$y_{\mathrm{est}, i}$.  The curve-fit coefficients are
given by the solution of the normal equations,
\begin{equation}
  \left(
    \begin{array}{ccc}
      \Sigma_0 & \Sigma_1 & \Sigma_2 \\
      \Sigma_1 & \Sigma_2 & \Sigma_3 \\
      \Sigma_2 & \Sigma_3 & \Sigma_4 
    \end{array}
  \right) 
  \left(
    \begin{array}{c}
      a_0 \\ a_1 \\ a_2
    \end{array}
  \right) =
  \left(
    \begin{array}{c}
      S_0 \\ S_1 \\ S_2
    \end{array}
  \right) ,
\end{equation}
where $\Sigma_n := \sum_{j=1}^{q} (x_j)^n$ and
$S_n := \sum_{j=1}^{q} y_j(x_j)^n$.
However, recomputing the sums in this system of equations
at each iteration is not computationally efficient.  
To make the on-the-fly
calculation of this curve fit feasible, we can instead
implement the coupled recurrence relations
\begin{equation}
  \setlength{\arraycolsep}{0ex}
  \renewcommand{\arraystretch}{1.4}
  \begin{array}{rcl}
    S_0^{(n)} &{}={}& S_0^{(n-1)} + y_{\mathrm{est}, n} + 
       y_{\mathrm{est}, n-m}\\
    S_1^{(n)} &{}={}& S_1^{(n-1)} - S_0^{(n-1)} + 
       m y_{\mathrm{est}, n-m}\\
    S_2^{(n)} &{}={}& S_2^{(n-1)} -2 S_1^{(n-1)} + S_0^{(n-1)} -
       m^2 y_{\mathrm{est}, n-m}
  \end{array}
\end{equation}
and then compute the fitted slope coefficient at time $t_n$ according to
\begin{equation}
  \begin{array}{l}
  \displaystyle
  a_1 = \frac{18(2q-1)}{q(q+1)(q+2)}\\ 
      \hspace{5mm}\displaystyle\times\left(
      S_0 + \frac{2(8q-11)}{3(q-1)(q-2)}S_1 + 
     \frac{10}{(q-1)(2q-1)}S_2\right).
  \end{array}
\end{equation}
For notational convenience, we have defined the times
corresponding to the last $q$ measurements to be
$t_n = 1-n$, so that the fitted slope at the current time
is simply $a_1$.  Thus, we can simply trigger the feedback
cooling algorithm on the sign of $a_1$, switching the potential
high or low for a positive or negative sign of $a_1$, respectively.
The speed of this algorithm is largely independent of the number of
samples used in the curve fit, assuming that the table of the
last $q$ values can be accessed quickly (i.e., it can fit into 
fast memory).

Later we will examine the effects of computation and signal propagation
delays on the cooling performance.  Such delays are easily accounted
for in the present algorithm by using the curve fit to extrapolate
forward in time.
To do this, we require the quadratic fit coefficient,
\begin{equation}
  \begin{array}{l}
  \displaystyle
  a_2 = \frac{30}{q(q+1)(q+2)}\\ 
      \hspace{5mm}\displaystyle\times\left(
      S_0 + \frac{6}{q-2}S_1 + 
     \frac{6}{(q-1)(q-2)}S_2\right),
  \end{array}
\end{equation}
so that we trigger on the value of the current-time slope,
\begin{equation}
  a_1 + 2 a_2 d,
\end{equation}
where $d$ is the delay in multiples of $\DtG$ from the time of the
most recent data used in the fit to the current time.
Due to the predictive nature of
this fitting scheme, we expect that the cooling should be robust
to small delays.

\noindent\textsl{\bfseries Gaussian Estimator.}
The other issue to be resolved is to maximize the efficiency of 
updating the Gaussian estimator.  One computationally expensive
aspect of the form of the equations (\ref{gest}) is the need to
evaluate transcendental functions.  A substantial speed improvement
results from introducing the variables
$x_1 := \sin(2\kt\Xe)$,
$x_2 := \cos(2\kt\Xe)$,
$x_3 := \Pe/\kt$,
$x_4 := 2\kt^2\Vxe$,
$x_5 := \exp(-2\kt^2\Vxe)$,
$x_6 := \Vpe/\kt^2$, and
$x_7 := \exp(-2\kt^2\Vxe)$,
we can rewrite the estimator equations in the form
\begin{equation}
  \setlength{\arraycolsep}{0ex}
  \renewcommand{\arraystretch}{1.4}
  \begin{array}{rcl}
    dx_1 &{}={}& (4\pi \kt^2 x_2 x_3 -4\eta\Gamma x_1^{\,3} x_4^{\,2} x_5^{\,2})dt\\
       && + \sqrt{8\eta\Gamma} x_1 x_2 x_4 x_5\dWe 
    \\ 
    dx_2 &{}={}& -(4\pi \kt^2 x_1 x_3 +4\eta\Gamma x_1^{\,2}x_2 x_4^{\,2} x_5^{\,2})dt\\
       && - \sqrt{8\eta\Gamma} x_1^{\,2} x_4 x_5\dWe 
    \\
    dx_3 &{}={}& -\Vmax x_1 x_5 dt + \sqrt{8\eta\Gamma}x_1x_5x_7\dWe
    \\
    dx_4 &{}={}& (8\pi \kt^2 x_7 -4\eta\Gamma x_1^{\,2} x_4^{\,2} x_5^{\,2})dt\\
       && + \sqrt{8\eta\Gamma} x_2x_4^{\,2} x_5\dWe 
    \\
    dx_5 &{}={}& [-8\pi \kt^2 x_5x_7 +4\eta\Gamma 
       (x_1^{\,2}x_5-x_2^{\,2}x_4^{\,2}) x_4^{\,2} x_5^{\,2}]dt\\
       && - \sqrt{8\eta\Gamma} x_2x_4^{\,2} x_5^{\,2}\dWe 
    \\
    dx_6 &{}={}& (-4\Vmax x_2x_5x_7 +\Gamma[1+x_5^{\,4}(1-2x_2^{\,2})]\\ 
       &&-8\eta\Gamma x_1^{\,2}x_5^{\,2}x_7^{\,2})dt\\
       && - \sqrt{2\eta\Gamma}[1-2( x_4+2x_7^{\,2}) x_5] x_2\dWe 
    \\
    dx_7 &{}={}& (2\pi \kt^2 x_6 -\Vmax x_2 x_4x_5 
       -4\eta\Gamma x_1^{\,2} x_4 x_5^{\,2} x_7)dt\\
       && + \sqrt{8\eta\Gamma} x_2 x_4 x_5 x_7\dWe ,
    \\ 
  \end{array}
  \label{fastgest}
\end{equation}
with
\begin{equation}
  \dWe = \frac{dr}{\sqrt{\eta\kapt}} + \sqrt{2\eta\Gamma}(1+x_2 x_5) dt ,
\end{equation}
thus obviating any need for transcendental-function evaluation.
However, it turns out that the evolution of these equations via
the Euler method is much less stable than the evolution of the
original equations.  This is due to the fact that there are pairs of 
variables [$(x_1, x_2)$ and $(x_4, x_5)$] that evolve separately
but must satisfy consistency conditions [$x_1^2+x_2^2 = 1$ and
$x_5 = \exp(-x_4)$]; also a large stochastic fluctuation can
push these values into unphysical ranges (e.g., $x_1>1$).
A better strategy is to emulate the results of an Euler solution
to the original estimator equations, and hence 
update $x_1$, $x_2$, and $x_5$ using the increments
\begin{equation}
  \setlength{\arraycolsep}{0ex}
  \renewcommand{\arraystretch}{1.4}
  \begin{array}{rcl}
    \Delta x_1 &{}={}& x_1[\cos(\twoktDeltaXe) - 1] + x_2\sin(\twoktDeltaXe)
    \\ 
    \Delta x_2 &{}={}& x_2[\cos(\twoktDeltaXe) - 1] - x_1\sin(\twoktDeltaXe)
    \\ 
    \Delta x_5 &{}={}& x_5\exp(-\Delta x_4),
  \end{array}
  \label{fastgestmod}
\end{equation}
where
\begin{equation}
  \twoktDeltaXe = 4\pi\kt^2 x_3\DtG + \sqrt{8\eta\Gamma}x_1x_4x_5
     \Delta W_\mathrm{e},
\end{equation}
$\Delta W_\mathrm{e}(t) := \int_{t}^{t+\DtG}\dWe$,
and $\Delta x_4$ is the Euler-update increment from 
Eqs.~(\ref{fastgest}).  These equations can then be expanded in powers
of $2\kt\Delta\Xe$ to avoid computing the transcendental functions.
A final trick to help stabilize the equations is to multiply the
$x_1$ and $x_2$ variables at each time step by the normalization quantity
$[1 - (x_1^{\,2} + x_2^{\,2})]/2$, which explicitly maintains the
identity $x_1^2+x_2^2 = 1$ without requiring the evaluation of a
square root.

\noindent\textsl{\bfseries Execution Times.}
To evaluate how quickly the control algorithm can be iterated, we 
implemented the control algorithm on a modern, general-purpose, serial
microprocessor (1.25 GHz Alpha EV68).  A single iteration of the
estimator according to Eqs.~(\ref{gest}) and a curve fit, including
all steps necessary to compute the control output, can be completed in
about 260 ns.  Switching to the faster method of Eqs.~(\ref{fastgest})
with the more stable updates (\ref{fastgestmod}) can be completed
in 140 ns with a truncation at 
fourth order in $\twoktDeltaXe$, and 120 ns with a second-order
truncation (including a normalization
step).  The cooling performance is essentially the same with all 
these algorithms, in that they all reach the final temperature;
the exception is an algorithm that uses Eqs.~(\ref{fastgest})
without the modifications of Eqs.~(\ref{fastgestmod}), which has
substantially worse cooling performance.  The tradeoff in 
the truncated-expansion algorithms
is that lower-order truncations are faster but also
more unstable and thus require more resets of the estimator.


Based on this performance, 
and assuming Moore's law, general-purpose hardware will not
be able to iterate this algorithm in 3 ns for about a decade.  
However, even presently available, specialized technologies
such as multicore processors and field-programmable gate arrays
(FPGAs) can likely be used to significantly decrease the execution time.
For example, because the curve fit can be moved onto a separate 
logical processor, the estimator can be updated more quickly
(84 ns for the fourth-order version, 70 ns for the second-order 
version); the extra 60 ns required to generate the feedback
value via the curve fit then represents an additional delay, but
does not limit the rate at which the estimator can be iterated.
Further parallelization should bring these times down 
substantially.

\begin{figure}[tb]
  \begin{center}
     \includegraphics[scale=0.41]{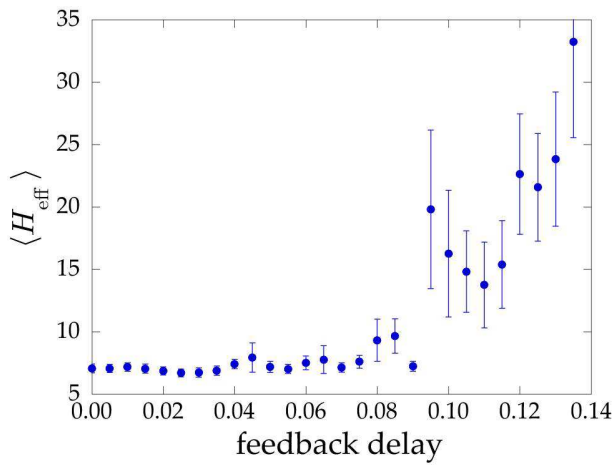}
  \end{center}
  \vspace{-5mm}
  \caption
        {(Color online) Final energies, measured over the interval from $t=90$
         to $100$, as a function of the feedback delay $\taud$.
         (Time is measured in units of the 5.8~$\mu$s 
         harmonic-oscillator period.)
         Cooling is according to the improved algorithm based
         on the Gaussian estimator, with extrapolation implemented
         in the curve fit; error bars reflect standard
         errors from averages over 128 trajectories.
         Other parameters are as in the canonical set.
	\label{fig:delaysweep}}
\end{figure}

\begin{figure}[tb]
  \begin{center}
     \includegraphics[scale=0.41]{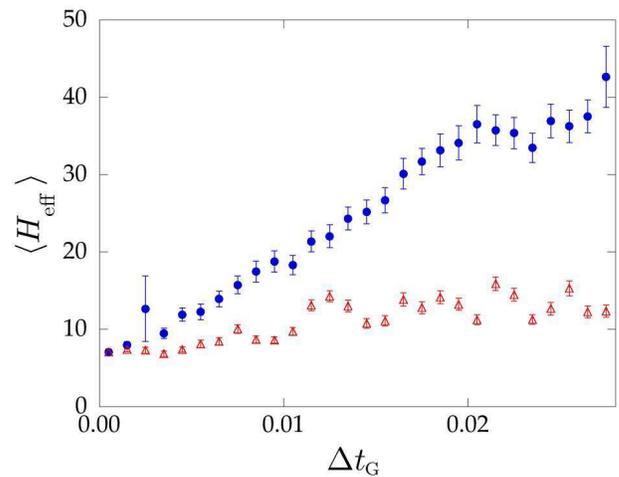}
  \end{center}
  \vspace{-5mm}
  \caption
        {(Color online) Final energies, measured over the interval from $t=90$
         to $100$, as a function of the Gaussian estimator
         time step $\DtG$. 
         (Time is measured in units of the 5.8~$\mu$s 
         harmonic-oscillator period.)
         Shown are results obtained by
         reducing the number of samples to keep the time interval
         for the quadratic curve constant (circles) as well as
         results obtained via the multiple-estimator staggering
         method to maintain a 300-point fit (triangles).
         Error bars reflect standard
         errors from averages over 128 trajectories.
         Other parameters are as in the canonical set.
	\label{fig:updatesweep}}
\end{figure}

\begin{figure}[tb]
  \begin{center}
     \includegraphics[scale=0.41]{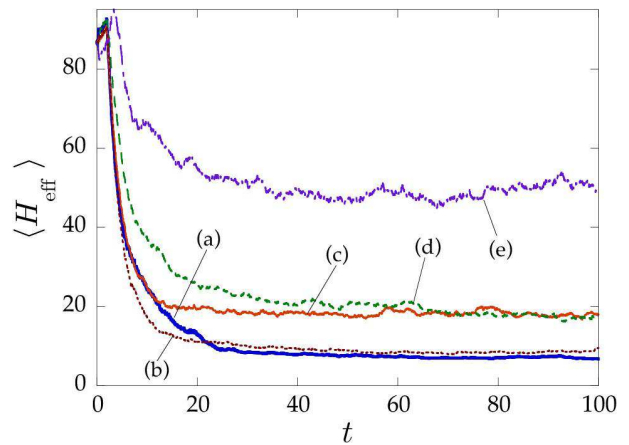}
  \end{center}
  \vspace{-5mm}
  \caption
        {(Color online) Ensemble energy evolution illustrating the combined effects of
         detection efficiency $\eta$, feedback delay $\taud$, 
         and estimator time step $\DtG$.  
         (Time is measured in units of the 5.8~$\mu$s 
         harmonic-oscillator period.)
         These parameter
         values for the curves (a)-(d)
         are given in the text; others are as in the
         canonical set.  Extrapolation is implemented in the
         curve fit, and multiple-estimator staggering is used
         to maintain a 300-point fit.
         Each curve represents an average over 128 trajectories.
	\label{fig:realistic}}
\end{figure}

\noindent\textsl{\bfseries Impact on Cooling Performance.}
Given these limitations in computing power it is necessary to 
evaluate how quickly the algorithm needs to be iterated in order
to achieve good cooling performance.  The cooling performance 
in the presence of feedback delay $\taud=d\DtG$ is shown in 
Fig.~\ref{fig:delaysweep}.  The cooling is very robust to the
delay, since we can compensate by extrapolating the curve fit.
The cooling behavior is essentially unaffected out to a delay
of about 460 ns, which is well above the 60 ns figure that we
mentioned for generating the feedback value, and thus additional
signal propagation and processing delays should not be too problematic.

Limitations on the time $\DtG$ required to iterate the Gaussian
estimator can also seriously impact the cooling performance.
Not only does a large $\DtG$ step reduce the quality of the 
estimated solution, but in fitting the quadratic curve over a 
finite interval, there are fewer points involved in the fit, 
reducing the effectiveness of the fit's noise reduction.
These effects on the cooling behavior are illustrated in
Fig.~\ref{fig:updatesweep}, which shows that the late-time ensemble
energies grow approximately linearly with $\DtG$.
Here, the 70 ns figure mentioned above for $\DtG$ yields a
final energy of around 25, compared to around 9 for the smallest
value of $\DtG=0.0005$ (3 ns) shown, a substantial decrease in
the cooling effectiveness.
One possible way to address this problem is to use multiple
Gaussian estimators that are staggered in time.  For example,
defining $\Delta t_\mathrm{min}=0.0005$, and given that the 
computational hardware is limited to $\DtG = N\Delta t_\mathrm{min}$
for some integer $N$, we can arrange to have $N$ independent
implementations of the Gaussian estimator executing in parallel.
They all receive the same measurement record as input, but they
are staggered in time such that one produces output at each 
small time step $\Delta t_\mathrm{min}=0.0005$.  In this way,
each of the individual solutions is of lower quality than
a fine-time-step solution, but the curve fit can be fit to
all of them (restoring a full 300 point curve fit), 
thus helping to stabilize the cooling algorithm.  The implementation
of this method is also displayed in Fig.~\ref{fig:updatesweep},
which shows that this method restores much of the
quality of the cooling that was lost in using the simpler
algorithm.  The rightmost point in this figure corresponds
to operating 55 parallel estimators.

So far, we have studied the performance impact on cooling of
each of the three equipment-limitation effects (detection
efficiency, feedback delay, and estimator time step) separately.
Some representative examples of their combined effect are shown
in Fig.~\ref{fig:realistic}.  The four curves, in order of
increasing departure from the ideal case, correspond to the 
following parameters:
(a) $\eta=1$, $\taud=0$, $\DtG=\Delta t_\mathrm{min}=0.0005$ (i.e., 3 ns in physical time); 
(b) $\eta=0.8$, $\taud=\DtG=10\Delta t_\mathrm{min}=0.005$ (i.e., 29 ns); 
(c) $\eta=0.7$, $\taud=\DtG=25\Delta t_\mathrm{min}=0.0125$ (i.e., 72 ns); 
(d) $\eta=0.5$, $\taud=\DtG=40\Delta t_\mathrm{min}=0.02$ (i.e., 116 ns); 
(e) $\eta=0.3$, $\taud=\DtG=50\Delta t_\mathrm{min}=0.025$ (i.e., 145 ns). 
The sums of the populations in the lowest two bands for these cases, 
computed between $t=90$ and $100$, are
$(94\pm 5)\%$, 
$(84\pm 5)\%$, 
$(45\pm 4)\%$, 
$(47\pm 4)\%$, and 
$(13\pm 2)\%$, 
respectively.  The cooling performance, even for the case that should
be within reach of current technology (d), is not unreasonable in
comparison to the ideal case (a), although cooling quickly becomes 
worse as these parameter values are relaxed (e).  Even with these
effects, this method is certainly suitable for long-time storage of
atoms in the optical potential and can still reach the ground
state with serviceable efficiency.




\begin{thebibliography}{74}
\bibitem[{not()}]{noteSME}
\bibinfo{note}{The terms in the stochastic master equation that describe
  homodyne detection are derived in~\cite{Wiseman93}. The term that describes
  spontaneous emission is given in, e.g.,~\cite{Jacobs96, Marte94}.}
\end{thebibliography}
\end{document}